\documentstyle[epsfig,multirow]{mn}

\newlength{\figurewidth}
\def\gtrsim{\ga}
\def\lesssim{\la}

\title[Nonlinear structure at Decoupling]{Can Nonlinear Structure Form 
at the Era of Decoupling?}

\author[N. J. Shaviv]{Nir J. Shaviv \\
Theoretical Astrophysics 130-33, California Institute of Technology, \\
Pasadena, CA 91125}

\begin{document}

\maketitle

\label{firstpage}

\begin{abstract}
The effects that large scale fluctuations had on small scale isothermal
modes at the epoch of recombination are analysed. We find that:
(a)~Albeit the fact that primordial fluctuations were at this epoch
still well in the linear regime, a significant nonlinear radiation
hydrodynamic interaction could have taken place. (b)~Short wavelength
isothermal fluctuations are unstable. Their growth rate is exponential
in the amplitude of the large scale fluctuations and is therefore very
sensitive to the initial conditions.  (c)~The observed CMBR
fluctuations are of order the limit above which the effect should be
significant.  Thus, according to their exact value, the effect may be
negligible or lead to structure formation out of isothermal
fluctuations within the period of recombination. (d)~If the
cosmological parameters are within the prescribed regime, the effect
should be detectable through induced deviations in the Planck
spectrum. (e)~The sensitivity of the effect to the initial conditions
provides a tool to set limits on various cosmological parameters with
emphasis on the type and amplitude of the primordial fluctuation
spectrum. (f)~Under proper conditions, the effect may be responsible
for the formation of sub-globular cluster sized objects at
particularly high red shifts. (g)~Under certain circumstances, it can
also affect horizon sized large scale structure.
\end{abstract}



\section{Introduction}

The problem of structure formation in the Universe has probably been
one of the foremost and most studied question in cosmology. Perhaps
the greatest achievement of cosmology was the prediction and following
discovery of the Cosmic Microwave Background Radiation -- the CMBR
(Penzias \& Wilson, 1965) and its anisotropy (Smoot et al. 1992). Not
only did it provide support for the Big Bang theory, it proved that
structure is a result of small fluctuations growing into large
inhomogeneities and not vice versa. The fact that the observed
fluctuations in the CMBR are small ($\Delta T/T \sim 10^{-5}$),
naively implies that one can treat the fluctuation within the linear
approximation when modeling the evolution of structure before the time
of last photon scattering, i.e., one can assume that fluctuation
modes of different wavelengths are completely decoupled from each
other before radiation-matter
decoupling.

The first to develop a linear theory for the perturbed Fridmann -
Robertson - Walker metric was Lifshitz (1946), and it can be found in
various text books such as Weinberg (1972), Peebles (1980) and Kolb \&
Turner (1990). The theory has since then been applied to describe
cosmologies with various components and various initial
parameters. Baryonic matter, radiation and other massless particles,
cold (massive) or hot (light) dark matter are some of the ingredients
that enter the primordial soup. While other parameters such as the
Hubble constant $H_0$, the vacuum energy through the cosmological
constant $\Lambda$, and the initial spectrum affect the evolution. A
review of various current cosmological models, their evolution and
implication can for example be found in White et al. (1994) with
emphasis on the microwave background radiation and in Primack
(1997). Less current reviews that cover the basic principles of
structure formation are found in the aforementioned textbooks.

One of the major parameters that affect the qualitative behaviour of
our universe and is highly relevant to the present paper, is the type
and form of the primordial spectrum of fluctuations.  The fluctuations
are generally classified into curvature (or adiabatic) and
isocurvature (or isothermal) fluctuations.  The first arise naturally
in inflationary scenarios (e.g., Liddle and Lyth, 1993, and
ref. therein). They are fluctuations of both the baryonic fluid and
the radiation and they propagate at the adiabatic speed of sound that
is equal to $c/\sqrt{3}$ if the radiation energy density dominates.
These waves are found to decay on scales smaller than the Silk scale,
namely, scales comparable to galaxy sized objects (Silk, 1967).
Therefore, a top-down structure formation is a natural consequence of
adiabatic perturbations. Small scale objects can form after
recombination (and initiate a bottom-up scenario) only if one adds the
undamped perturbation of a cold component as is the case in the
standard CDM model (Peebles 1982, Blumenthal et al. 1984 and Davis et
al. 1985) and its variants.  Isocurvature (or isothermal) fluctuations
on the other hand, are a natural consequence of topological defects
formed in phase transitions (e.g, strings, monopoles and textures) or
if more than one field contributes significantly to the energy density
during inflation. They correspond to fluctuations that alter the
entropy density but not the energy density. Unlike the first type of
fluctuations, these do not suffer from Silk damping. Consequently,
objects with a mass as small the post-recombination Jeans scale,
namely the size of globular clusters, can form after decoupling.

It is interesting and important to note that even if the primordial
spectrum was purely adiabatic, isothermal perturbations of a given
wavelength are formed as the second order effect of Purcell clustering
from adiabatic waves of similar wavelengths (Press \& Vishniac, 1979).
Through this effect, noninteracting particles (baryons) that are
viscously coupled to a stochastically oscillating background gas
(radiation fluid) undergo secular clustering. Originally, it was
thought that the effect can produce a bottom-up scenario even from a
pure baryon model with an adiabatic spectrum. However, the typical
post-recombination Jeans scale amplitude of the isothermal waves will
only be $\rho_{iso}/\rho\sim (10^{-2} - 1) \times \left( \delta
\rho_{ad} / \rho\right)^2$, with $\delta \rho_{ad} /\rho$ the typical
adiabatic fluctuations at horizon crossing. Namely, if the primordial
spectrum is flat and adiabatic, the typical isothermal fluctuation at
recombination is roughly $\delta \rho_{iso} /\rho \sim
10^{-10}-10^{-8}$. For a flat isothermal primordial spectrum, one
should expect typical amplitudes of $\rho_{iso}/\rho
\sim 10^{-5} - 10^{-4}$.

The smallness of nonlinear effects such as the Purcell clustering (or
shock waves which are of an even higher order) led to the consensus
that the evolution of the fluctuations can be treated linearly and
modes of different wavelengths are decoupled from each other.  This
delays the nonlinear treatment to the time when radiation does not
play any dynamic role anymore. At face value, it certainly appears to
be the case as $\delta\rho /\rho$, $v/c$, and $\delta T/T$ are all
much smaller than unity.  One should nevertheless be extremely careful
when assuming linearity, especially in view of the fact that not all
of the dimensionless parameters are actually smaller than unity.

One of the dimensionless numbers that appears in the solution of the
nonlinear fluctuations' equations of motion and that is not small at
all is the root of the radiation to gas pressure ratio.  Just before
recombination one finds that $\sqrt{P_{rad}/P_{gas}}\approx{10}^{5}$!
Consequently, we expect that the interaction between the radiation and
matter at this period will have profound impact on the evolution of
the fluctuations.

In this paper we examine the linear hypothesis and its validity by
adding the force large scale perturbations exert on short wavelength
waves. We begin in \S2 by overviewing the problem of solving the
nonlinear equations of motion and estimating the effect with a very
simple analysis.  We proceed in \S3 to write the Newtonian equations
describing the evolution of short wavelength isothermal modes. In \S4
we analyse the simplified solution, while in \S5 we proceed to
estimate the effect in a few cosmological scenarios. In \S6 \& \S7, we
study the possible ramifications to structure formation and study the
possibility of measuring and using this effect in the study of
cosmological parameters. In \S8 we show that the effect can influence
large scale structure as well.


\section{The Effect}

 At the lowest order of approximation, waves of different wavelengths
 are decoupled from each other. This suggests that a small scale
 isothermal wave will, at this order, witness an isotropic and
 homogeneous medium around it. However, at the next order of
 approximation, one has to include the small perturbations (of order
 $\delta \rho / \rho \sim 10^{-4}$) of all the other scales in the
 vicinity of the isothermal wave.

One can simplify the problem by assuming that the isothermal wave is
localised to a region of size $d$ that is at least several times the
wave's wavelength (such that the wave parameters are well
defined). The scale $d$ separates between two types of perturbations:
those with a wavelength $\lambda < d$ and those with $\lambda > d$.
The contribution for example to the radiative force of the former
group vanishes in the lowest order as the spatial correlation between
the wave and perturbation yields a net vanishing result.  The latter
group can contribute a net effect only if the temporal correlation
between the perturbation and the isothermal wave does not vanish as
well; namely, it will do so only if the perturbation does not have
enough time to oscillate rapidly inside the region under
consideration.  This implies that for a given period $\Delta t$, the
contribution of waves with $\omega \Delta t \gtrsim 1$ averages out
leaving no net contribution. The only waves that have a net
contribution are those for which $\omega \Delta t \lesssim 1$; that
is, if $d$ is chosen smaller than the wavelengths of waves with
$\omega \sim 1/\Delta t$, its exact value is unimportant.  Hence, we
solve for short wavelength isothermal waves in a region where the long
wavelength perturbations can be considered constant and
homogeneous. We cannot however, assume that the local environment is
isotropic.


We assume for simplicity that the large scale perturbations are
optically thick adiabatic waves of the form:
\begin{equation}
{\delta \rho_{a} \over \rho} = \delta_a \sin \left( {\bf k}_a
\cdot {\bf x}
- \omega_a t\right),
\end{equation}
where $\delta \rho_{a}$ is the perturbation of the matter density. It
is proportional to the temperature perturbations:
\begin{equation}
{3 \delta T_{a} / T} = {\delta \rho_{a} / \rho} ,
\end{equation}
namely, the fluctuations in the radiation field and in the matter are
synchronised.  The sole force responsible for accelerating the matter
is the radiative force. If observed from the baryon rest frame, the
force is due to a net radiative flux originating from an anisotropy in
the radiation field.  This anisotropy or shear between the radiation
and matter fluids can be found through the baryon acceleration.
 
Using the adiabatic speed of sound $v_{(a)}$ to relate ${\bf k}_a$
with $\omega_a$ and the continuity equation to relate $\bf{v}_a$ with
$\delta \rho_a$, the velocity of the baryonic fluid in the adiabatic
wave is found to be:
\def\na{{\bf{\hat n}_a}}
\def\ni{{\bf{\hat n}_i}} 
\begin{equation}
{\bf v}_a = \delta v_{(a)}\sin \left( {\bf k_a}\cdot
{\bf x} - \omega_a t\right) \na,
\end{equation}
with $\na$ being a unit bector in ${\bf k_a}$'s direction.  The
acceleration or the force per unit mass is simply $d{\bf v}_a/dt$,
consequently, if the opacity per unit mass is $\kappa$, the net
radiative flux $H$ is given by:
\begin{equation}
{\bf  H} = {{\bf a_a}\over \kappa} = {1\over \kappa} {d {\bf 
v}_a\over d t} =-{ \delta_a v_{(a)} \omega_a \over \kappa} \cos
\left( {\bf k_a}\cdot {\bf x} - \omega_a t\right) \na .
\label{H_k_eq}
\label{eq_4}
\end{equation}

 The form for an isothermal wave travelling in a small region
 perturbed by the wave above is:
\begin{equation}
{\delta \rho_i \over \rho} = \delta_i \sin \left(
{\bf k_i}\cdot {\bf x} - \omega_i t\right).
\end{equation}
with $\delta T_i=0$ as it is isothermal. It will witness a constant
radiative flux ${\bf H}$ given by eq.~\ref{eq_4} as it is optically
thin.  However, if $\kappa$ is a function of density, the opacity and
therefore the force per unit mass, become a function of the wave's
phase. The predominant source of opacity before and during
recombination is Thomson scattering off the free
electrons. Consequently, the only period during which isothermal waves
vary the opacity is during recombination, when the number of free
electrons is proportional to ${\rho}^{-1/2}$.  Hence, the total
acceleration is:
\begin{equation}
{\bf a} = \kappa {\bf H} = \kappa_0 {\bf H}+ \delta \kappa_i {\bf H} 
\equiv {\bf a}_a + {\bf a}_i = {\bf a}_a
\left(1 - {1\over 2} {\delta \rho_{iso} \over \rho} \right).
\label{eq_5}
\end{equation}
The first term results with the large scale acceleration already
accounted for in the large scale fluctuation. The second term however,
incites a force that varies synchronously with the small scale
isothermal waves. It leads to an instability.

The total power input per unit mass by the latter term into the wave
is:
\begin{eqnarray}
p_{rad} & = &{\bf a}_i \cdot {\bf v}_i =  {1\over 2} v_{(a)} \delta_a 
\omega_a \cos \left( {\bf
k_a}\cdot {\bf x} - \omega_a t\right) \na \nonumber \\ & & \cdot~
\delta_i \sin
\left(
{\bf k_i}\cdot {\bf x} - \omega_i t\right) v_i \ni \nonumber \\ & = & 
 {1\over 2} v_{(a)} \omega_a \delta_a \cos \left( {\bf
k_a}\cdot {\bf x} - \omega_a t\right)  \nonumber \\ & & \times 
\sin \left(
{\bf k_i}\cdot {\bf x} - \omega_i t\right)^2 v_{(i)} \delta_i^2 \cos 
\theta, 
\end{eqnarray}
where $\ni$ denotes a unit vector in ${\bf k}_i$'s direction and
$\theta$ is the angle between $\ni$ and $\na$.  Again, we related
${\bf v}_i$ to $\rho_i$ through the continuity equation and the
isothermal speed of sound $v_{(i)}$.  When averaging the force over
time one has to recall that the large scale perturbations are assumed
not to vary over the relevant period - the duration of
recombination. Now we see why. Oscillating large scale perturbations
will average to a zero net contribution. Those that oscillate on a
longer time scale can be assumed constant:
\begin{equation}
\delta_a \cos \left( {\bf
k}\cdot {\bf x} - \omega t\right) \rightarrow \delta_\phi.
\end{equation}
Using $\left< \cos^2 \right> = 1/2$ to average the isothermal
oscillations, we find:
\begin{equation}
\left< p_{rad}\right> =  {1\over 4} v_a \omega \delta_\phi v_{(i)} 
\delta_i^2 \cos \theta.
\end{equation}
The total energy per unit mass in the acoustic wave is
\begin{equation} e = {1\over 2}
\delta_i^2 v_{(i)}^2/2.
\end{equation}
The $e$-fold growth rate of the wave's
amplitude is therefore:
\begin{equation}
r = {\left< p\right> \over 2 e} = {{1\over 4} v_a \omega \delta_\phi 
v_{(i)} \delta_i^2  \over 2 {1\over 2} \delta_i^2 v_{(i)}^2} \cos 
\theta=
{1\over 4} \omega \delta_\phi {v_{(a)} \over  v_{(i)}} \cos \theta.
\end{equation}
Let the duration of recombination be $\Delta t$. The growth factor G, 
or the number of $e$-folds the wave will grow if traveling in the 
direction of the large scale shear (i.e., if $\theta=0$), is:
\begin{equation}
G = r \Delta t = {\delta_\phi\over 4}  {v_{(a)} \over  v_{(i)}} 
\left( \omega \Delta t \right) \equiv G_0 {\delta_\phi \over 
10^{-4}}~{\rm with} ~G_0 = {\cal O}(1),~
\label{eq_14}
\label{G_simp}
\end{equation}
where we have inserted the ratio of the speeds of sound at
recombination assuming radiation dominance over baryonic matter. We
have also taken $\omega \Delta t \sim 1$, as waves with a larger
$\omega$ average out. $G_0$ is a constant of order unity which cannot
be found in this approximate estimate.  $\delta_\phi$ at recombination
will have a distribution of values. If however, the typical value is
of order $10^{-4}$, then the typical growth rate is $G_0$
$e$-folds. Namely, we expect that the nonlinear effect is at least as
important as the linear evolution of the isothermal wave.  It is now
also evident why isothermal waves are unstable while adiabatic are
mostly stable. One only has to replace $v_{(i)}$ by $v_{(a)}$ to find
that the adiabatic waves are unstable only for $\delta_{\phi}$ of
order unity and not much less.

Eq. \ref{G_simp} is useful as an estimate, but a more general
expression for the growth parameter $G$ should not include an
assumption on the type, speed, or optical depth of the large scale
waves, but instead use directly the radiation flux observed in the
baryon rest frame.  By using eq. \ref{H_k_eq}
and letting $\kappa H$ vary during recombination, we have:
\begin{equation}
G \sim {1\over 4 v_{(i)}} \int_0^{\Delta t} \kappa H dt.
\label{more_acc_eq}
\end{equation}
Note that $H$ is the flux observed in the baryon rest frame.


\section{Newtonian equations of motion}

The general nonlinear relativistic equations describing the evolution
of the perturbations are extremely difficult, if not impossible, to be
solved analytically.  We therefore simplify the problem until it
becomes amenable to an analytical treatment without losing the basic
physical ingredients.  It is generally not possible to decouple the
system into different wavelengths and solve each one
separately. Nonetheless, if the nonlinearity is small, namely, the
interaction between modes of different wavelengths is small, we can
write the equations describing each mode as decoupled equations to
which a coupling correction is added. Moreover, we have seen in \S2
that the coupling is essentially an effect large modes (adiabatic or
isothermal) have on small isothermal waves.  That is, the large
horizon sized modes are largely unaffected and can be integrated as
uncoupled modes, whereas the small modes are mostly affected only by
the large modes and not by themselves.  Consequently, the short
wavelengths modes can be solved independently as the equation for each
one is the basic uncoupled linear equation to which a linear source
term is added. This term depends only on the history of the long
wavelengths modes. Actually, this statement is true only as long as
the amount of energy transferred into the short wavelengths is
negligible when compared to the energy in the larger scale. This will
be further discussed in \S8.

To summarise, the assumptions made in order to simplify the equations 
are:
\begin{itemize}
 \item The short wavelength fluctuations are pure isothermal waves -
 i.e., there are no perturbations to the radiation field.
\item The
 fluctuations are of a wavelength much smaller than the Horizon size
 and are therefore solved under the Newtonian approximation.
\item
 The short wavelength fluctuations are affected only by large
 wavelength modes.
\item
 The evolution of the large wavelength modes is decoupled and is
 assumed to have been solved a priori.
\item
 The short wavelength modes are solved in  localised regions where
 the effect of the large scales can be considered as homogeneous.
\end{itemize}

\begin{figure}
\centerline{\epsfig{file=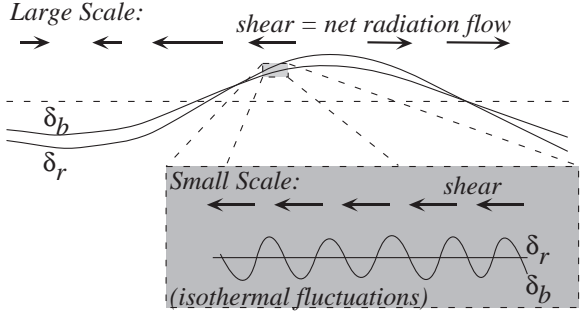,width=3in}}
\caption{
The physical system we solve. The equations describe the evolution of
small scale perturbations embedded in an environment that is
inhomogeneous on large scales and can have both a perturbation in the
baryon density $\delta_b$ and in the radiation energy density
$\delta_r$.  The large scale inhomogeneities are the Horizon scale
perturbations. The small scales are very short wavelength isothermal
perturbations solved locally in a small region where the large scale
can be considered uniform, or in other words, that the large
scale flux is spatially constant. This approximation is possible
because the small scale wavelengths considered are much smaller than
the inhomogeneities affecting them.  }
\label{system_figure}
\end{figure}

The system solved is described in figure \ref{system_figure}.

We start by writing the equations of motion describing Newtonian {\em
isothermal} fluctuations.  The simplest equations describing matter
fluctuations of scales much smaller than the horizon size and
velocities much smaller than the speed of light are the Newtonian
equations of motion, corrected to include the expansion of the
universe. These are (see for example Kolb \& Turner 1990):
\begin{eqnarray}
0& = & \dot{\rho}_1 +  3 {\dot{a}\over a} \rho_1 + {\dot{a}\over a}  
({\bf
r}\cdot \nabla)\rho_1 + \rho_0 \nabla \cdot {\bf v}_1
\nonumber \\ 0&=&
\dot{\bf v}_1 +  {\dot{a}\over a} {\bf v}_1 + {\dot{a}\over a}  ({\bf
r}\cdot \nabla){\bf v}_1 + {v_{s}^2 \over \rho_0} \nabla \rho_1 + \nabla
\phi_1 - {\bf f}_{\rm rad}  \nonumber \\ 0&=&
\nabla^2 \phi_1  - 4\pi G \rho_1 ,
\end{eqnarray}
where the various quantities have their usual meaning. Here $v_{s}$
is the isothermal speed of sound.  The last term in the second
equation is the force per unit mass acting on the baryonic fluid.  It
can be expressed with the help of the flux ${\bf H}$ observed in the
baryonic rest frame and the opacity per unit mass $\kappa$. It can
also be expressed with the radiation ``velocity'' ${\bf v}_r$ -- the
velocity with which the baryons need to move in order for them not to
observe a net radiation flow, and a characteristic time
$\tau_{e\gamma}$ for the transfer of momentum between the
electron-proton fluid and the photons. If the angular dependence of
the photon energy density distribution $\delta \rho_r(\theta, x) $ is
expanded into Legendre polynomials:
\begin{equation}
{\delta \rho_r(\theta, x, t) \over \rho_r (x, t)} = \sum_{\ell 
=0}^{\infty} \left(2 \ell + 1\right) P_{\ell} (\cos \theta) 
\sigma_\ell (x,t), 
\end{equation}
then the radiation ``velocity'' is given by $v_r \equiv \sigma_1 / 4$
and the flux observed in the baryon rest frame is $\left(\sigma_1 - 4
v_m\right) c \rho_r$ with $v_m$ being the material velocity. Either 
way, the
total radiative force per unit mass $f_{rad}$ can be written as:
\begin{equation}
{\bf f}_{rad} = {\bf H} \kappa = {{\bf v}_m - {\bf v}_r \over 
\tau_{e\gamma}}.
\end{equation}
We will work with the latter expression that uses the effective
velocity of the radiation fluid $v_r$ as it is easier to work with
when coupling the baryons to the radiation.  For nonrelativistic
velocities and Thomson scattering, the term can be written as:
\begin{equation}
{\bf f}_{rad} = {4\over 3} \sigma_T a T^4 {{\bf v}_m - {\bf v}_r 
\over c} {n_e \over 
\rho_b}
\equiv { {\bf v}_m - {\bf v}_r \over \tau _{e\gamma}}.
\end{equation} 
$n_{e}$ is the number of free electrons, $a$ denotes the radiation
density constant, and $\sigma_T$ is the Thomson cross section.

Normally when linearised and Fourier transformed, this term which
describes the radiation drag has a contribution only from the wave for
which the linearized equations of motion are written.  However, the
contribution of waves of all scales should be taken into account if
through the nonlinear interaction they produce a correction
synchronised with the waves. When linearising, we find the following
terms:
\begin{equation}
{\bf f}_{rad} = {{{\bf v}_m}_1 - {{\bf v}_r}_1 \over
{\tau_{e\gamma}}_{0}} - {{{\bf v}_m}_0 - {{\bf v}_r}_0 \over
{\tau_{e\gamma}}_{0}} {{\tau_{e\gamma}}_{1}\over  
{\tau_{e\gamma}}_{0}}.
\end{equation}
The first term is responsible for the radiation drag due to the
difference between the wave's material velocity and the radiation
``velocity''.  The second term is due to the nonlinear effect that all
other waves induce on this particular one through the change in the
medium's opacity {\em and} the shear between the material velocity and
the effective radiation velocity of all other waves. The value of the
large scale shear ${\bf v}_{m0}-{\bf v}_{r0}$ is the value at the
region in which the small scale wave is solved for. The latter
contribution is synchronised with the wave through the changes in the
opacity. It is of second order and it should be integrated over
time. Clearly, the contribution of waves oscillating at frequencies
different than the particular wave   averages to zero, thus, the
only net contribution arise from waves that do not oscillate over
the relevant integration time; that is to say, only the shear of the
large scale waves  contributes. If a wave oscillates $n$-times
during the period, its net contribution will on average fall as
$n^{-1}$. When integrated numerically in \S5, the contribution of
smaller scales is taken into account automatically and their
contribution indeed falls with $\omega$.

The radiation field is not perturbed for optically thin, isothermal
fluctuations. Consequently, ${{\bf v}_r}_1$ vanishes.
If we define the large scale shear as:
\begin{equation}
\Delta {\bf v}_0 \equiv {{\bf v}_m}_0-{{\bf v}_r}_0,
\end{equation}
and write
\begin{equation}
{\tau_{e\gamma}}_1 =  {\tau_{e\gamma}}_0 \left(\alpha {\rho_1 \over
\rho_0} + \beta {T_1 \over T_0}\right), 
\end{equation}
then the radiation term is:
\begin{equation}
{\bf f}_{\rm rad} = { {\bf v}_1 \over {\tau _{e\gamma}}_0} + { 
\Delta{\bf
v}_0 \over {\tau _{e\gamma}}_0} { {\tau _{e\gamma}}_1 \over {\tau
_{e\gamma}}_0 } \equiv { {\bf v}_1 \over {\tau _{e\gamma}}_0} + { 
\Delta{\bf
v}_0 \over {\tau _{e\gamma}}_0}  
\left(\alpha {\rho_1\over \rho_0}+ \beta {T_1\over T_0}\right).~~~
\end{equation}
The second term does not vanish if the opacity is perturbed by the
isothermal waves. By definition, the temperature isn't perturbed and
$T_1=0$. If however the opacity is a function of the density, then
$\alpha \neq 0$ and a new term is added to the Newtonian equations of
motion. The opacity is predominantly Thomson scattering, consequently,
$\tau$ is proportional to the number of free electrons.  The number of
free electrons varies only during recombination. At this time $n_{e}$
is inversely proportional to the root of the density. To be more
exact, the fraction of ionised protons at equilibrium $X_{eq}$ is
given by Saha's equation:
\begin{equation}
{1-X_{eq}\over X_{eq}^2} = {4 \sqrt 2 \zeta(3) \over \sqrt{\pi}} \eta
\left( T\over m_e\right)^{3/2} \exp(E_{ion} / T),
\end{equation}
with the baryon to photon ratio $\eta$ being proportional to the
density and $E_{ion}$ the ionisation energy of hydrogen. Through
differentiation we find:
\begin{equation}
\alpha = {\partial \ln X_{eq} \over \partial \ln \rho} = - {1-X_{eq}. 
\over 2-X_{eq}}  
\end{equation}
That is, before recombination commences, $X_{eq}$ is approximately
unity and the parameter $\alpha$ vanishes. Afterwards it will grow in
size and reach the asymptotic value of $-1/2$.  Note that for the
above equation to be valid, we need to assume that the time scale for
ionisation equilibrium ($e+p \longleftrightarrow H+\gamma$) is shorter
than the dynamic time scale.  This implies that among other
assumptions, the equations are valid only before freeze-in which takes
place not too long after the advent of recombination.  We have also
neglected Helium ionisation and its influence on the density
dependence of the opacity. However, the low abundance of Helium
results with a much smaller effective $\alpha$, and together with the
larger speed of sound, they both reduce the effect during the period
of Helium recombination.

Define $\delta =
\rho_1/\rho_0$ and spatially expand $\delta$, ${\bf v_1}$ and $\phi_1$
in Fourier integrals proportional to $\exp\left( -i {{\bf k}\cdot r
/ a(t)}\right)$. The result is: 
\begin{eqnarray}
0&=& \dot\delta_k - {i {\bf k}\over a}\cdot {\bf v}_k \nonumber \\
0&=& a \dot{\bf v}_k + \dot{a} {\bf v}_k - i {\bf k} v_{s}^2 \delta_k - 
i
{\bf k} \phi_k + a { {\bf v}_1 \over {\tau _{e\gamma}}_0} + \alpha a {
\Delta{\bf v}_0 \over {\tau _{e\gamma}}_0} \delta_k ~~ \nonumber \\
0&=& \phi_k + {4 \pi G \rho_0\over k^2} a^2 \delta_k . 
\end{eqnarray}
The new phenomena discussed here arise from the introduction of the
last term in the second equation. The term is proportional to the
density perturbations (through opacity changes) and to the integrated
large scale shear $\Delta v_0$ which varies in space.  Even with the
additional radiation term, the irrotational flow is still decoupled
from the rotational part. On the other hand, the rotational part is
affected from the irrotational one. This interesting effect can
introduce a seed magnetic field from any isothermal fluctuation
present, as is described in Shaviv \& Levin (1998).

By taking the irrotational part of the second equation (through the
dot product with ${\bf k}$), and using the first and last equations,
we find that $\delta_k$ obeys the equation:

 \[
 \ddot{\delta}_k + \left( {2 \dot{a}\over a} + {1\over
 \tau_{e\gamma}} \right) \hskip -1pt \dot{\delta}_k + \left( {k^2
 v_{s}^2} - 4 \pi G \rho_0 - i \alpha {\Delta \hskip-1pt {\bf v}_0
 \hskip -2pt \cdot\hskip -2pt {\bf k}\over
 \tau_{e\gamma} }\right) \hskip -1pt
 \delta_k \hskip -1pt = \hskip -1pt 0
 \]
\begin{equation} 
\label{simpeq}
\end{equation}

This equation (and its derivation) should be compared with the
standard derivation of the equations describing isothermal waves when
all nonlinear interactions are neglected (e.g., Coles and Lucchin,
1995).  The solution of the linear equation consists of two modes. One
describes a highly damped wave with a damping time scale of
$\tau\sim\tau_{e\gamma}$. The other describes a frozen mode. Although
the latter is formally unstable, the large radiation drag effectively
reduces the growth rate to a value much smaller than the universe
expansion rate.

\section{Approximate Solution}
The general solution to eq.~\ref{simpeq} depends on the exact time
dependence of $a$, $\dot{a}$, $\rho_0$, $\Delta v_0$ \&
$\tau_{e\gamma}$. We are however, interested in the evolution over the
short period of decoupling, a period that is much smaller than the age
of the Universe at the time. Moreover, we are particularly interested
in circumstances were the growth rates are large, in fact, larger than
the rate in which the various parameters change during
recombination. This will allow us to assume that the above variables
are quasi-static.

The system described by eq. \ref{simpeq} has five different time
scales of which some depend on the wave number. The first two are the
expansion age of the universe and the related Jeans timescale. For a
flat matter dominated universe, the first is:
\begin{equation}
\tau_{\rm u}\equiv {3 a \over  2 \dot{a}},
\end{equation}
while the second is:
\begin{equation}
\tau_j \equiv 1/\sqrt{4\pi G \rho_0} = \sqrt{3/2} \tau_u.
\end{equation}
The last equivalence assumed a flat matter dominated universe.
The third timescale is the isothermal oscillation time:
\begin{equation}
\tau_{o}^{-1} \equiv { |k| v_{s}}.
\end{equation}
The fourth is the electron-photon relaxation time $\tau_{e\gamma}$
that describes how fast baryon and 
radiation exchange energy.
The newly introduced timescale is the time associated with the 
radiation
inhomogeneity:
\begin{equation}
\tau_{rad}^{-1} \equiv \sqrt{\alpha {\Delta {\bf v}_0 \cdot {\bf k} / 
\tau_{e\gamma} } }.
\end{equation}

\begin{figure*}
\centerline{\epsfig{file=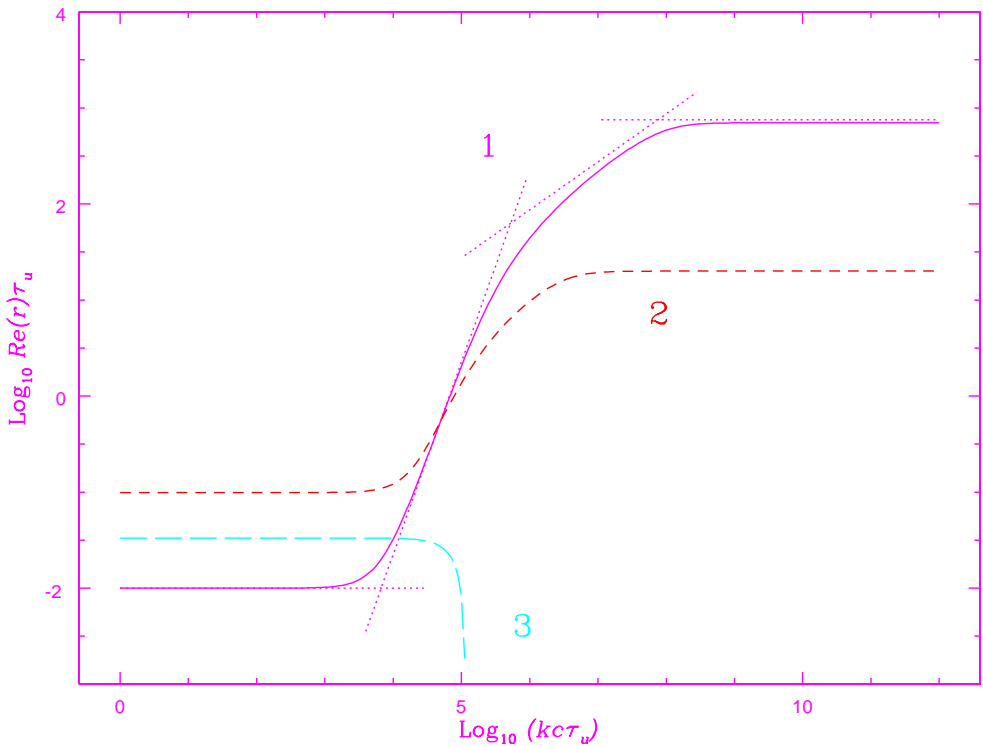,width=\figurewidth,angle=-90}}
\caption{
The growth rate as a function of the wave number depicted for three
examples.  In all three examples shown $v_s = 10^{-5} c$ and
$\alpha=1/2$.  In example 1: $\Delta v_0 = 3\times 10^{-4}$ and
$\tau_{e\gamma}=0.01 \tau_u$, in example 2: $\Delta v_0 = 1\times
10^{-4}$ and $\tau_{e\gamma}=0.1 \tau_u$, while in example 3: $\Delta
v_0 = 1\times 10^{-5}$ and $\tau_{e\gamma}=1/30 \tau_u$.  The first
two examples correspond to case A of table 1 while the third example
for which $\delta v_0 < v_s/\alpha $ corresponds to case B in the
table. Only in the former two cases do we find that the growth rate is
increased tremendously for large values of $k$.  The curve of example
1 has the approximate behaviour corresponding to the four different
regions of case A marked with small dotted lines in addition to the
exact value of the root.  In the four different regions, we have $k
\propto const, k^2, k^{1/2}, const $. They do not exist in example 3,
and while they do exist in the second example, the behaviour is not as
clear as in the first one. }
\label{rer_figure}
\end{figure*}

We will assume here that the electron-photon relaxation time is much
smaller than the universe's expansion time scale. The reason being
that when the timescale surpasses the expansion scale, the solution is
in any case invalid because the timescale for establishing ionisation
equilibrium becomes very long.  The $\dot{a}/a$ term can therefore be
neglected. The time scales themselves depend on the cosmic time as
well, however, since we are interested in timescales shorter than both
the expansion of the universe and the timescale for recombination of
hydrogen, we can for simplicity assume them to be constant or
quasi-stationary. In such a case, the equation's solution is
exponential with the roots:
\begin{equation}
r = -{1\over 2 \tau_{e\gamma}} \pm \sqrt{{1\over 4 \tau_{e\gamma}^2}-
\left({1\over \tau_{o}^2} - {1\over \tau_{j}^2} - i
{1\over \tau_{rad}^2}\right)}.
\label{roots}
\end{equation}
The waves' growth rate is given by the real part of $r$, namely
$\Re(r)$.  It will be governed by the largest rate (or smallest time
scale) in the system. The oscillatory behaviour is given by the
Imaginary part -- $\Im(r)$. For an optically thick universe with
$\tau_{e\gamma} <\tau_u$, we find two general cases. The two cases and
their various regions of $k$ in which the rates behave differently,
are summarised in table
\ref{cases_table} and exemplified in figure \ref{rer_figure} where a
few sample graphs of $\Re(r(k))$ are shown.

\begin{table*}
\noindent
\begin{tabular}{|c c | c c|}
\hline \hline
\multicolumn{4}{|c|}{Case A: $\Delta v_0 > v_s/\alpha$} \\\hline
\hline
\multicolumn{2}{|c|}{Boundary between regions} &
\multicolumn{2}{c|}{Growth Rate: $\Re(r)$ }   \\ \hline
 & & \multirow{2}{30mm}{\centerline{$k \ll k_{-\infty}$}} &
 \multirow{2}{62mm}{$\Re(r) \approx {\tau_{e\gamma}/\tau_{j}^2}$} \\
\cline{1-2}
 \multirow{2}{22mm}{$\tau_{rad}^2=\tau_j\tau_{e\gamma}$} &
 \multirow{2}{33mm}{$k_{-\infty}\equiv {1\over \alpha \Delta {\bf v}_0
 \tau_{j}}$} & & \\
\cline{3-4}
 & & \multirow{2}{30mm}{\centerline{$k_{-\infty}\ll k \ll k_{0}$}} &
 \multirow{2}{62mm}{$\Re(r) \approx \tau_{e\gamma}^3 / \tau_{rad}^4 =
 \tau_{e\gamma} \left( \alpha \Delta {\bf v}_0 \cdot {\bf k}\right)^2
 $}\\
\cline{1-2}
 \multirow{2}{20mm}{$\tau_{e\gamma}=\tau_{rad}$} &
 \multirow{2}{33mm}{$k_{0}\equiv {1\over \alpha \Delta {\bf v}_0
 \tau_{e\gamma}} $} & & \\
\cline{3-4}
 & & \multirow{2}{30mm}{\centerline{$k_{0} \ll k \ll k_{\infty}$}} &
 \multirow{2}{62mm}{$\Re(r) \approx \pm {1\over \sqrt{2} |\tau_{rad}|}
 = \pm \sqrt{ {\alpha \over 2} {\Delta {\bf v}_0 \cdot {\bf k} \over
 \tau_{e\gamma}}} $ } \\
\cline{1-2}
 \multirow{2}{20mm}{$\tau_{rad}=\tau_o$} &
 \multirow{2}{33mm}{$k_{\infty}\equiv {\Delta {\bf v}_0 \over v_s }{
 \alpha \over v_s \tau_{e\gamma}}$} & & \\
\cline{3-4}
 & & \multirow{2}{30mm}{\centerline{$k_{\infty} \ll k$}} &
 \multirow{2}{62mm}{$\Re(r) \approx \tau_o / (2 \tau_{rad}^2)= {\alpha
 \over 2} {\Delta {\bf v}_0 \cdot \hat{\bf n} \over v_s} {1\over
 \tau_{e\gamma}}$}\\
\cline{1-2}
 & & & \\
 \hline  
%
\hline 
\multicolumn{4}{|c|}{Case B: $\Delta v_0 < v_s/\alpha$} \\\hline
\hline
\multicolumn{2}{|c|}{Boundary between regions} &
\multicolumn{2}{c|}{Growth Rate: $\Re(r)$ }   \\ \hline
 & & \multirow{2}{30mm}{\centerline{$k \ll k_{1}$}} &
 \multirow{2}{62mm}{$\Re(r) \approx {\tau_{e\gamma}/\tau_{j}^2}$} \\
\cline{1-2}
\multirow{2}{20mm}{$\tau_{o}=\tau_u$} & 
\multirow{2}{35mm}{$k_{1}\equiv {1/ (\tau_u  v_s)}$} &  &  \\
\cline{3-4}
 & & \multirow{2}{30mm}{\centerline{$k_{1}\ll k \ll k_{2}$}}  &  
\multirow{2}{62mm}{$\Re(r) \approx - \tau_{e\gamma} / \tau_o^2 = - 
\tau_{e\gamma} v_s^2 k^2 $}\\
\cline{1-2}
\multirow{2}{20mm}{$\tau_{e\gamma}=\tau_o$} & \multirow{2}{35mm}{$k_2 
\equiv {1 /(\tau_{e\gamma} v_s)}$}  & & \\
\cline{3-4}
 & & \multirow{2}{30mm}{\centerline{$k_{2} \ll k$}}  &  
\multirow{2}{62mm}{$\Re(r) \approx - 1/2\tau_{e\gamma} $}\\
\cline{1-2}
 & & & \\
\hline
\end{tabular}

\caption{
The growth rate of the small scale waves for various wavelengths
under the two relevant cases of ${\Delta v_0} > v_s/\alpha$ and
${\Delta v_0} < v_s/\alpha$, with $\tau_{e\gamma}<\tau_u$ ( i.e., as
long as the universe is optically thick). Note that only if
$\Re(r)>\tau_u^{-1}$ is the solution appropriate. Otherwise, instead
of an exponential solution, the system will evolve through a power law
behaviour with a typical growth scale of $\tau\sim\Re(r)$.  }
\label{cases_table}
\end{table*}


When analysing eq.~\ref{roots}, one finds that it admits two general 
forms of behaviour depending on   whether $\Delta v_0$ is smaller or
larger than $v_s /\alpha$. The main difference between the two cases 
is that large $k$ waves in the former case are damped while in the 
latter they are unstable and amplify. More specifically, one has:

\noindent
{\bf Case A: $\Delta v_0 < v_s /\alpha$} \\ \noindent When the shear
velocity between the radiation and baryon fluid is smaller than the
critical value, the shear cannot overcome the radiation drag.  Even
though the quantitative analysis may change, the general behaviour is
the one found when the shear velocity is altogether neglected. That is
to say, long wavelengths for which $\tau_o > \tau_u$ are frozen in the
plasma. Although they do have a positive growth rate, it is small when
compared with the growth rate of the horizon. Waves with smaller
wavelengths for which $\tau_o < \tau_u$ are damped, and when
$\tau_{e\gamma} \sim \tau_o$, the damping rate is saturated at its
asymptotic value of $1/2\tau_{e\gamma}$.

\noindent
{\bf Case B: $\Delta v_0 > v_s /\alpha$} \\ \noindent The behaviour of
the system is radically different when the large scale shear velocity
is larger than the critical value.  At large wavelengths, the waves
are still frozen in and the behaviour is the same as in the previous
case. On the other hand, at shorter wavelengths the waves' growth rate
is dominated by $\tau_{rad}$.  Instead of being damped or frozen-in,
the waves can in this case be unstable. The growth rate of waves
propagating in the direction of the shear increases with $k$ and it
saturates at small wavelengths when acoustic oscillations become
important.  The maximum or asymptotic growth rate is obtained for
values of $k$ larger than:
\begin{equation}
k_\infty ={\Delta {\bf v}_0 \over v_s }{
 \alpha \over v_s \tau_{e\gamma}},
\end{equation}
i.e., for scales smaller than the Jeans scale by a factor of
order ${\tau_u/\tau_{e\gamma}}$. The typical values of
$\tau_u/\tau_{e\gamma}$ when the rate is large are of order 10,
implying that the typical scales would be 10 times smaller than the
Jeans scale. The maximum growth rate is then given by:
\begin{equation}
\Re(r) \approx \tau_o / (2 \tau_{rad}^2)= {\alpha
 \over 2} {\Delta {\bf v}_0 \cdot \hat{\bf n} \over v_s} {1\over
 \tau_{e\gamma}}.
\label{max_growth}
\end{equation}
It is now evident why isothermal waves are crucial. The speed of sound
$v_s$ in isothermal waves is much smaller than the speed of light,
therefore, the typical shears that cause nonlinear interactions are
of order ${\cal O}(v_s/c)$ and not ${\cal O}(1)$! An appreciable 
nonlinear
effect can take place already at a dimensionless amplitude of
$\delta\rho/\rho \sim 10^{-5}$ or $10^{-4}$!

We are interested in the integrated effect of the growth rate in the
time interval between a given $t_0$ and a given later time $t_1$.
Hence, we define the following growth parameter:
\begin{equation}
G(k,\hat{\bf n}) = \int_{t_0}^{t_1} \Re(r(k,\hat{\bf n})) dt.
\label{growthint}
\end{equation}
If the rate given by eq.~\ref{max_growth} is larger than all other
inverse timescales of the system on which the value of the rate can
change (i.e., both the expansion and the recombination timescales)
then the growth factor of the isothermal waves is approximately given
by:
\begin{equation}
{\delta_{t_1}(k,\hat{\bf n})\over \delta_{t_0}} \approx \exp(
G(k, \hat{\bf n})).
\label{growth_eq}
\end{equation}
The recombination time $\Delta t \equiv t_1-t_0$ is the shortest
timescale on which the growth rate changes. As a consequence, the
condition for the approximation to hold can be written as $\Re(r) \gg
\Delta t^{-1}$, and it can be translated through eq.~\ref{growthint}
to $G \gg 1$.

We proceed to investigate the rate of growth in two ways:

\begin{itemize}
 \item The nonlinear growth rate equation (eq.~\ref{growth_eq}) is
 averaged over the distribution of perturbation amplitudes and
 direction of motion in order to get the average growth of the
 amplitude of the waves.

 \item We explore systems that reach nonlinear amplitudes by examining
 the volume fraction that does develop nonlinearities and the possible
 effects triggered by these nonlinear regions.
\end{itemize}

\subsection{Average Growth}
We now proceed to estimate the {\em average} growth of amplitude of
small scale waves for which $k \gtrsim k_{\infty}$.  Although the
physical scale is small enough not to be {\it directly} detectable in
the near future, these modes have the fastest growth rate and are
therefore most interesting.

The averaging procedure of the amplification  described by equation
\ref{growth_eq} should be divided into two separate averages over two
distributions. First, the expression should be averaged over the
various possible propagation directions since unlike the nonlinear
case, the dispersion relation is now a function of the direction of
${\bf k}$ (with respect to the direction of the large scale shear) and
not only its absolute value. Second, the growth function is found for
a given local value of the shear, however, it is certainly not linear
in its amplitude and one should therefore average over the
distribution of large scale shears.

We first average expression \ref{growth_eq} over the $4\pi$ possible
directions of $\hat{\bf n}$ (the direction of ${\bf k}$) while keeping
in mind that $G$ is linearly proportional $\hat{\bf n}
\cdot \hat{\bf n}_0$. We get:
\begin{equation}
\left\langle {\delta_{t_1}(k,\hat{\bf n})\over \delta_{t_0}}
\right\rangle_{\hat{\bf n}}  = \int d\hat{\bf n} 
{\delta_{t_1}(\hat{\bf
n})\over \delta_{t_0}} \end{equation}
\[
 \phantom{
\left\langle {\delta_{t_1}(k,\hat{\bf n})\over \delta_{t_0}}
\right\rangle_{\hat{\bf n}}
}=  \int d\hat{\bf n} \exp( G_\infty(k) (\hat{\bf
n}\cdot \hat{\bf n}_0)) =  {\sinh(G_\infty(k))\over G_\infty(k)}
\]
with $G_\infty(k)=G(k,\hat{\bf n} = \hat{\bf n}_0)$, i.e., its maximum
value.

Next, we proceed to average over the distribution of shears
$\Delta{\bf v}_0$. This distribution will clearly depend on the form
of the density distribution. The natural choice for the latter is a
Gaussian distribution. Although it does not necessarily have to be the
case, it is the theoretically most reasonable distribution expected
for the density fluctuations\footnote{Although it is a natural
consequence of inflationary scenarios, it is markably different for
models in which cosmic defects provide the origin of the fluctuations
(e.g. Turok 1996).}.  The distribution of the absolute value of the
shears is similar in form to the distribution of the absolute
velocities calculated from density perturbations in large scale
structure (e.g. Coles and Lucchin \S18.3), that is, if the
distribution for the density fluctuations is Gaussian then it will be
Guassian for the components of the shear vector in each axis, and the
distribution for the absolute value will be given by a Maxwellian
distribution. If the root mean square of $G_{\infty}$ is $G_{rms}$,
the distribution itself is:
\begin{equation}
P(G_\infty) = \sqrt{54\over\pi} {G_\infty^2\over G_{rms}^3} \exp 
\left[- {3\over 2} \left( G_\infty\over G_{rms}\right)^2 \right].
\label{eq_gdistr}
\end{equation}

The average perturbation growth factor is therefore:
\begin{equation}
\left\langle {\delta_{t_1}\over \delta_{t_0}}
\right\rangle_{all} \end{equation}
\[ ~~~~~ = \int_{0}^\infty \sqrt{54\over\pi} {G_\infty^2\over 
G_{rms}^3} \exp 
\left[- {3\over 2} \left( G_\infty\over G_{rms}\right)^2 \right]
{\sinh(G_\infty)\over
G_\infty} dG_\infty \]
\[ ~~~~~ = \exp \left( G_{rms}^2 \over 6 \right) \equiv {\cal 
A}(G_{rms}). \]

The function ${\cal A}(G_{rms})$ describes the amplification factor of
the fluctuations.  Clearly, for $G_{rms}$ much less than or of order
unity, the effect is small or negligible. Moreover, the numerical
values obtained, such as a 10\% increase for $G=1$, are only
approximate since the required condition $G \gg 1$ for the validity of
eq.~\ref{growth_eq} is not fulfilled. Nonetheless, already at
$G_{rms}=5$, small scale isothermal fluctuations will amplify by
almost 2 orders of magnitude. They will grow by 5 orders of magnitude
for $G_{rms}=8$ and for $G_{rms}=12$ the growth will be a staggering
10 orders of magnitude. Note that if the growth is saturated when the
amplitude reaches nonlinearity then the average growth is meaningful
only if the amplified amplitude is less than unity. Moreover, unlike
the case when $G_{rms} \sim 1$, the solution here is much more
accurate because regions that contribute the most to the growth are
those in the high $G_\infty$ tail, where the approximate solution is
valid. For example, the average growth when $G_{rms}=8$ is of 5 orders
of magnitude and the main contribution originates from $G_\infty
\sim \ln(10^5) \approx 11.5 \gg 1$.

Also worth mentioning is that in the averaging procedure over the
fluctuation distribution, we are essentially averaging an exponential
function $\sim \exp (G_\infty)$; thus, if the distribution for the
amplitudes does not fall fast enough for large values of $G_\infty$
(which is proportional to $\delta \rho/ \rho$), the averaging integral
diverges. Such a divergence occurs if the high $\delta \rho/ \rho$
tail is for example a power law. The implication is that there will
always be a finite volume (i.e., volumes that are not exponentially
small) where nonlinearity can be reached if the distribution has a
wide tail. A wide tail can in this case be any distribution with a
power law tail or even a distribution with an exponential tail,
provided the tail of the distribution is in the form $\propto \exp(
-G_{\infty}/G_0)$ with $G_0>1$.


\subsection{The nonlinear regions}

The large exponential growth associated with small regions where the
shear is large, is of high importance. In these regions that are small
for a modest value for $G_{rms}$, the growth may be large enough as to
amplify waves out of the linear regime and let the nonlinear evolution
of these wave commence. Consequently, it is interesting to estimate
what fraction of space will actually be filled with nonlinear waves by
the end of recombination.

\begin{figure*}
\centerline{\epsfig{file=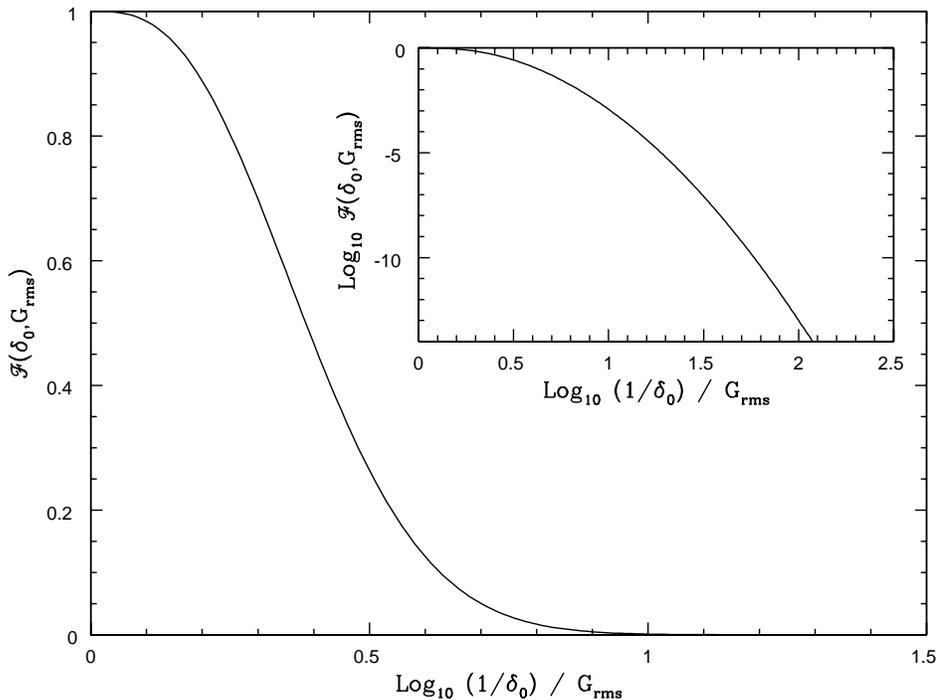,width=\figurewidth,angle=-90}}
\caption{
The fraction of regions that reach nonlinearity -- ${\cal
F}(\delta_0,G_{rms})$ as a function of $Log_{10} (1/\delta_0) /
G_{rms}$.  Any fraction less than $\sim 10^{-6}$ will be theoretically
impossible to detect and correlate with gradients in the CMBR at a
$3\sigma$ level, as there are only about $10^7$ statistically
independent regions for which the shear is substantially different and
uncorrelated.}
\label{saturfunc}
\end{figure*}

The largest growth factor in a given region is given by
$\sim\exp(G_\infty)$.  Thus, if the typical small scale fluctuation
amplitude before amplification is initially $\delta_0$, then only if
$\delta_0 \exp(G_\infty) \gtrsim 1$ will there be waves that grow out
of linearity, i.e., only if $G_\infty\gtrsim - \ln \delta_0$. If the
density fluctuations are Gaussian, the distribution of growth factors
$G_\infty$ is Maxwellian and given by eq. \ref{eq_gdistr}. The
fraction of space in which there are waves that reach nonlinearity is
then given by
\[
{\cal F}(\delta_0, G_{rms})  =  \hskip -2pt \int_{- \ln \delta_0}^{\infty} 
\sqrt{54\over\pi} {G_\infty^2 \over G_{rms}^3} \exp 
\left[- {3\over 2} \left( G_\infty\over G_{rms}\right)^2 
\right]dG_{\infty}
\]
\begin{equation}
~~ = {\rm
erfc} \left( - {3\over 2} { \ln \delta_0 \over G_{rms}}\right)  
-\sqrt{6\over\pi} \exp\left( -{3\over 2}\left( \ln \delta_0
\over G_{rms}\right)^{2}\right) { \ln \delta_0 \over G_{rms}}.~
\end{equation}
Here erfc denotes the complementary error function.
The function $\cal F$ is plotted in figure
\ref{saturfunc}. As an example, for $G_{rms} \sim 5$, $\delta_0$ 
should be roughly $10^{-4}$ to have 1\% {\it of the volume} reach
nonlinearity.

If we could observationally detect all the regions in the
background microwave sky that reach nonlinearity (see e.g. \S5) then
the absolute number of regions showing nonlinearity will be given by
the fraction that does develop it multiplied by the total number of
statistically independent regions. The latter number is roughly $4\pi
\ell^2$ with $\ell$ the multipole order corresponding to the typical
shear producing wavelengths. By taking a co-moving $k\sim 0.1 Mpc^{-1}
$ as the typical correlation length for the shear producing zones (see
e.g. \S 5 or fig.~\ref{runput3_figure}) we get:
\begin{equation}
\ell \approx (6000~h^{-1}Mpc)k \sim 1000,
\end{equation}
namely, to get a statistically significant number of regions, one 
must have:
\begin{equation}
{\cal F}(\delta_0,G_{rms}) \gg {1\over 4 \pi \ell^2} \sim 10^{-7}.
\end{equation}
Evidently, for our example of $G\approx 5$, any initial isothermal
perturbations that has $\delta_0 \gtrsim 10^{-6}$ is theoretically
detectable and will produce at least 100 different nonlinear regions.
If the primordial spectrum of fluctuations is isothermal and flat, and
the expectation for $\delta_0$ is $\sim 10^{-4}$ then for any
$G_{rms}>3$ one should in future be able to detect the effect. For
$G\gtrsim 10$ more than 50\% of the universe reaches nonlinearity at
$z\sim 1000$.  In the case where the isothermal wave amplitude has the
smallest plausible value, namely when  the primordial spectrum is flat and
adiabatic and the Press-Vishniac (1979) effect produces isothermal
waves with an amplitude $\delta_0 \sim 10^{10}-10^{-8}$, the effect
will be detectable for any $G_{rms}\gtrsim 7.5$.

\section{Estimating the effect in actual cosmological scenarios}

 The largest attainable growth parameter is roughly:
\begin{equation}
G_\infty \approx  {\alpha \over 2} {{\Delta v_0} \over v_s} 
{\Delta t \over \tau_{e\gamma}}.
\end{equation}
where $\Delta t$ is the approximate duration of decoupling.  $v_s$ is
the isothermal sound speed which at recombination is about 4.8 orders
of magnitude less than the speed of light. Therefore, when
recombination commences and $\tau_{e\gamma}\sim 10^{-2} \tau_u$ (for
an approximate duration of $\Delta T\approx 10^{-1} \tau_u$), the
shear velocity should be about $3\times 10^{-5} c$ to get a 2 order of
magnitude increase and $6\times 10^{-5} c$ to get as much as a 10
orders of magnitude increase.  As recombination continues, the shear
increases while $1/\tau_{e\gamma}$ and the optical depth decrease
roughly by the same factor, such that the rate of growth is
effectively unchanged during this period, even though the opacity
changes by more than two orders of magnitude.

The interesting question that will ultimately determine the fate of
the small scale fluctuations is therefore a quantitative one -- what
is the exact value (i.e. the exact r.m.s.) of the shear exerted by
large scales and what is the integrated rate equal to?

To answer the quantitative question, a numerical simulation for the
behaviour of the large scale modes was carried out.  The numerical
code is a modification of the COSMICS code developed by Bertschinger
and Ma. The code was used to integrate the linearised equations of
general relativity, matter and radiation in the synchronous guage, and
subsequently normalize  the power spectra to the COBE measured
quadrupole moment. A description of the equations solved and the
numerical algorithm used can be found in Bertschinger (1995) and Ma
\& Bertschinger (1997). With the modification, the maximum growth rate
given by eq.~\ref{max_growth} is calculated and integrated to give the
growth parameter.

The evolution of each co-moving large scale wave vector $k$ is
followed over the course of the particular scenario's cosmological
history. For each $k$ integrated, the code calculates the
evolution of the wave amplitude of the various species in the
primordial soup, including among other, the angular moments of the
photon distribution $\Delta_\ell(k,t)$. These are to be precise, the
Fourier and Legendre decomposition of the photon brightness
temperature perturbation $\Delta({\bf x},{\bf {\hat n}},t) \equiv
\Delta T/T$ of the photons traveling in the direction of ${\bf \hat
n}$; more specifically, they are defined through:
\begin{equation}
 \Delta({\bf x},{\bf {\hat n}},t) \hskip -1pt = \hskip -3pt \int d^3k e^{i {\bf 
k}\cdot {\bf
 x}} \sum_{\ell=0}^{\infty}(-1)^\ell (2\ell+1) \Delta_\ell({\bf k},t)
 P_\ell({\bf \hat k} \cdot {\bf \hat n}).~
\end{equation}
The program also integrates the appropriate unnormalized growth factor:
\begin{equation}
g_\infty(k,t) = \int_0^t {\alpha \over 2} {{\Delta v_0} \over v_s} 
{dt \over \tau_{e\gamma}},
\label{g_eq}
\end{equation} 
where $\Delta v_0$ is the unnormalized shear.

To normalize both $\Delta_\ell(k,t)$ and $g_\infty(k,t)$, one should
take into account the original, primordial power spectrum $P(k)=A
k^{4-n}$. When $n=1$ the power index corresponds to a flat
Harrison-Zel'dovich power spectrum for which the total power per
decade is constant.  We normalize the spectrum to fit the measured
COBE quadrupole moment. This is achieved through the separation of the
angular correlation function $C(\theta)$ into its moments:
\begin{equation}
C(\theta) = \sum_\ell {2 \ell +1 \over 4\pi} C_\ell 
P_{\ell}(\cos\theta),
\end{equation}
with $P_\ell$ the $\ell$'th Legendre polynomial, and writing the
moments as a function of the normalized photon power distribution
functions:
\begin{equation}
C_\ell = 4\pi \int d^3 k P(k) \Delta_\ell^2(k).
\label{cl_eq}
\end{equation}
The quadrupole moment is related to the second moment $C_2$ of the
angular correlation function by the definition:
\begin{equation}
Q_{rms-PS} \equiv T_0 \left( 5 C_2 \over 4 \pi\right)^{1/2},
\end{equation}
through which the constant $A$ of the power spectrum can be
normalised.  The value used for the quadrupole moment is
$Q_{rms-PS}=18\mu K$ \footnote{The four year $1-\sigma$ COBE DMR
results are $Q_{rms-PS}=18\pm 1.6~\mu K$ if $n$ is assumed to be 1 (a
flat spectrum). If it is not constrained, then
$Q_{rms-PS}=15.3^{+3.6}_{-2.6}~\mu K$ (Bennet et al. 1996). The final
value of $G_{rms}$ is proportional to the value of $Q_{rms-PS}$
used.}.  The r.m.s. of $G_\infty$ can then be related to the
unnormalized growth rate (eq. \ref{g_eq}) through the
expression\footnote{The coefficient before the integral (unity) is
chosen to be consistent with $P(k)$ defined to be the total power
spectrum, thus consistent with eq.~\ref{cl_eq} as well.}:
\begin{equation}
G_{rms}^2 = \int d^3k P(k) g_\infty^2(k). 
\end{equation}

\begin{figure*}
\centerline{\epsfig{file=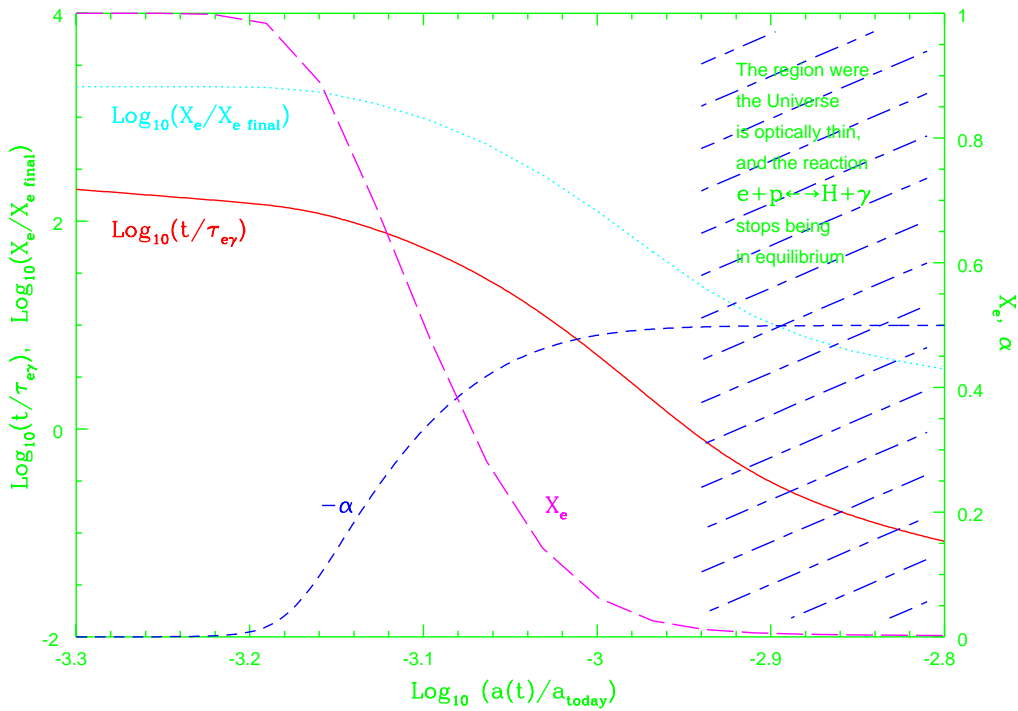,width=\figurewidth,angle=-90}}
\caption{The various physical variables that affect the net shear in 
the system and its effect on small scales as a function of expansion
radius of the universe for a sCDM model with
$H_0=66~km~s^{-1}~Mpc^{-1}$ and $\Omega_b=0.05$. The variables are
$\alpha = \partial \log X_e / \partial \log \rho$ which directly
affects the strength of the force induced on small scales, $X_e$, the
number of free electrons to which the total force is proportional,
$t/\tau_{e\gamma}$ gives the mean free path of an electron in the
photon ``fluid'' to horizon scale ratio, while $X_e / X_{e~final}$ is
the ionisation to residual ionisation ratio. If it is much greater
than unity, the reaction $e + p \longleftrightarrow H+\gamma$ is still
in equilibrium and the solution is valid .}
\label{runput4_figure}
\end{figure*}

\begin{figure*}
\centerline{\epsfig{file=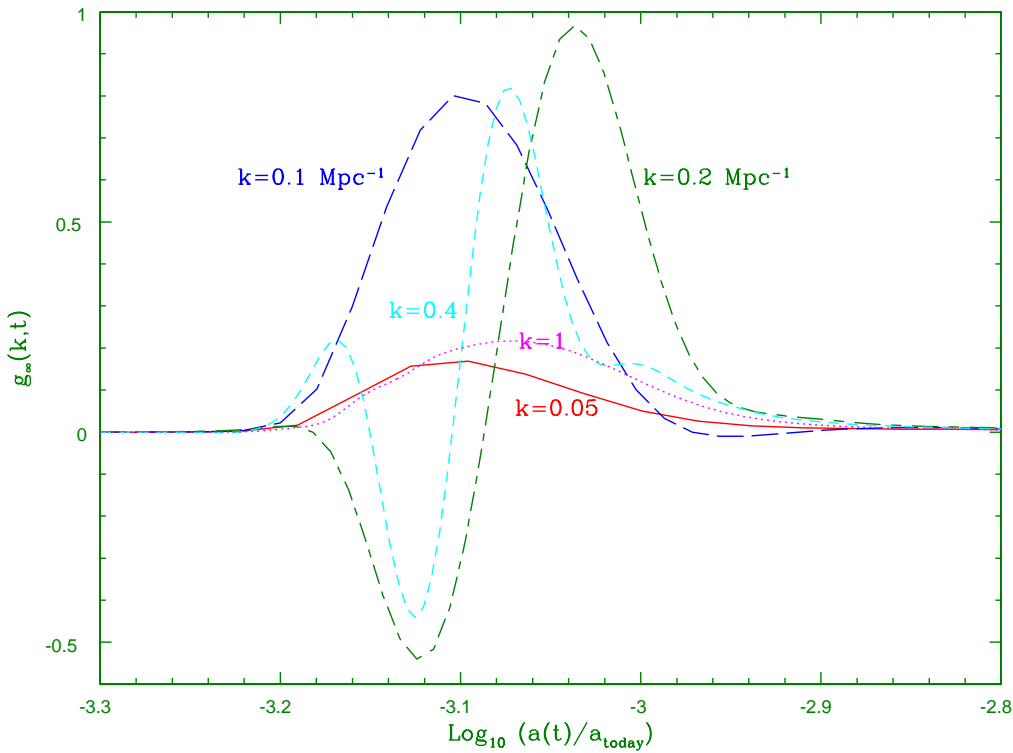,width=\figurewidth,angle=-90}}
\caption{
The (unnormalized) evolution of the instability growth rate
$g_\infty(k,t)$ as a function of the universe expansion scale for five
different wave numbers; namely, the contribution to the large scale
shear from various wavelengths during the time of recombination. For
sub-Horizon scales which are not too small, waves with larger $k$
oscillate more times during recombination and cancel out some of their
contribution. However, when $k$ is increased further, the optical
depth corresponding to one wavelength falls below unity and the
radiation fluid and baryon fluid stop oscillating synchronously. As
the baryon fluid loses its radiation support, its oscillations
decreases, as can be seen in the figure.  As a consequence, the
contribution from large wavenumbers does not fall as fast as $1/k$, as
is evident in fig. 8. The scenario is a standard CDM with
$H_0=66~km~s^{-1}~Mpc^{-1}$, $\Omega_b=0.05$ and a flat adiabatic
primordial spectrum (run 1). }
\label{runput1_figure}
\end{figure*}

\begin{figure*}
\centerline{\epsfig{file=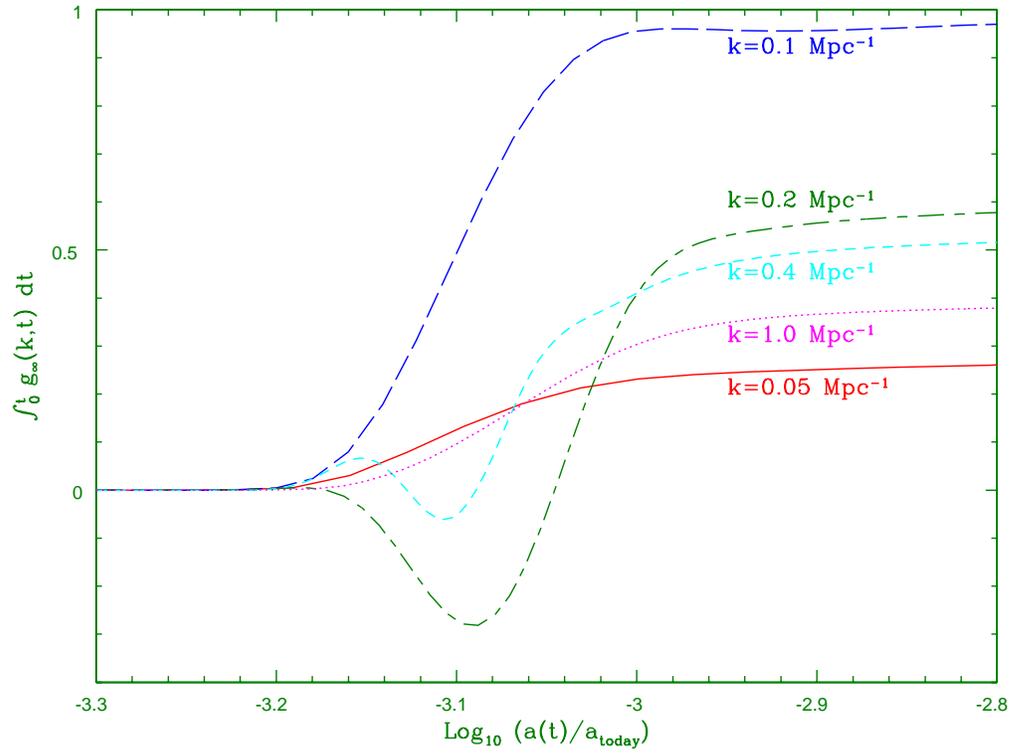,width=\figurewidth,angle=-90}}
\caption{
The same as the previous figure except that plotted are the time 
integrated
growth rates.
}
\label{runput2_figure}
\end{figure*}

\begin{figure*}
\centerline{\epsfig{file=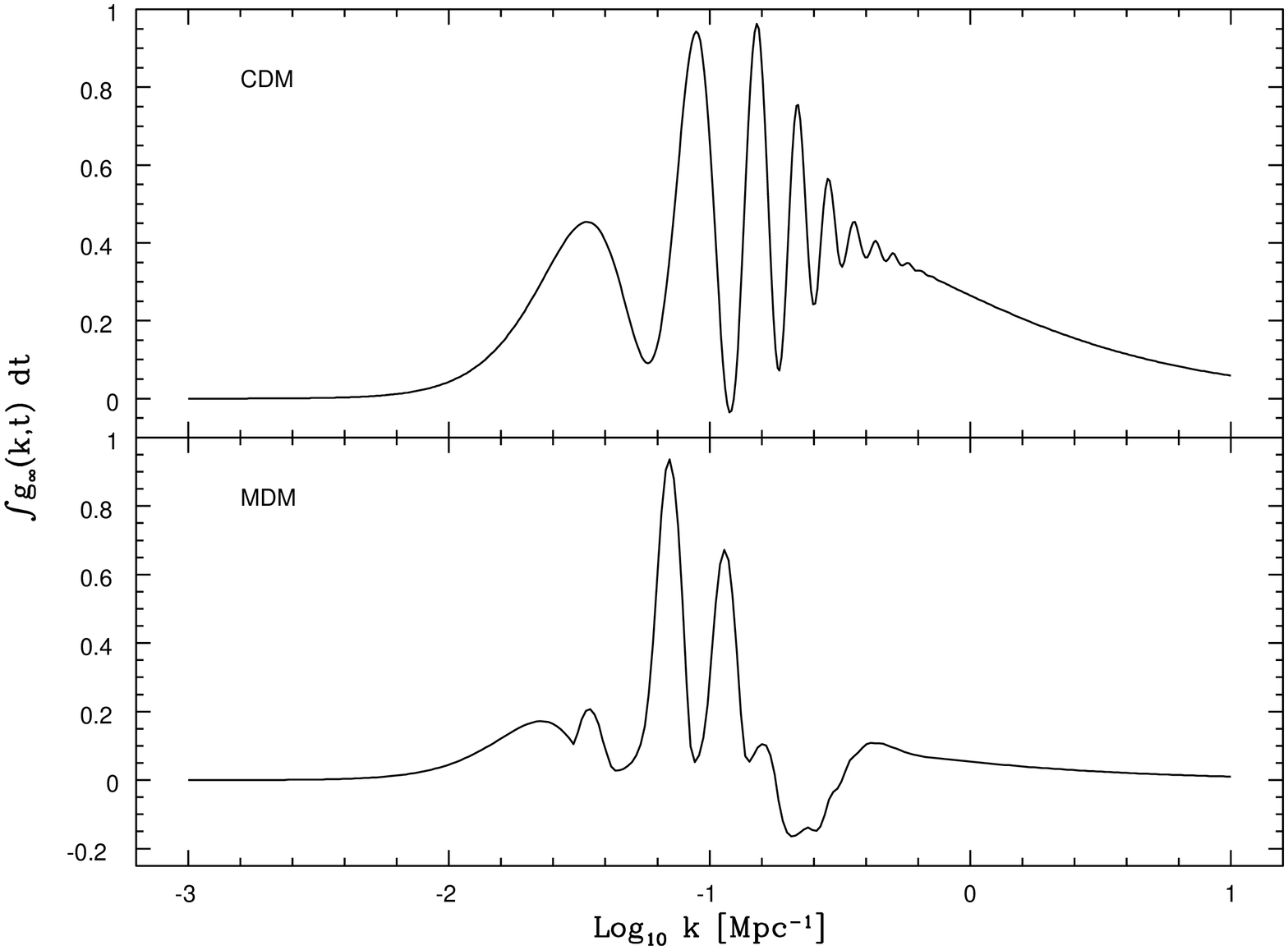,width=\figurewidth,angle=-90}}
\caption{
The (unnormalized) integrated growth rate as a function the co-moving
wavenumber $k$ for two cases. One is the standard CDM (run 1) while
the other is a mixed Hot-Cold dark matter (run 4), with parameters
given in table 2.  }
\label{runput3_figure}
\end{figure*}

A few cosmological scenarios were checked, all of which correspond to 
flat universes. Typical results are summarised in figures
\ref{runput4_figure}-\ref{runput3_figure}. In figure
\ref{runput4_figure} we find the various physical parameters that
affect the large scale shear. Figures \ref{runput1_figure} and
\ref{runput2_figure} depict the unnormalized growth rate and its
integrated value (the growth parameter) as a function of time for
several co-moving wavenumbers in a standard CDM model (run 1). The
last figure describes the integrated growth parameter as a function of
co-moving wavenumber for two scenarios - a standard CDM and a mixed
Hot-Cold dark matter scenario. We find that most of the contribution
comes from a co-moving $k\sim 0.1~Mpc^{-1}$ but a significant
contribution arises from scales that are smaller or larger by as much
as an order of magnitude (or even more in the case of CDM).

The scenarios simulated are summarised in table \ref{result_table}. We
find that the typical r.m.s. of $G_\infty$ varies from slightly less
than unity to more than 5, i.e., the average growth of the small
scale isothermal waves in flat cosmologies can range from 10\%
to 2 orders of magnitude, depending on the particular scenario.

\begin{table}

{\parindent 0pt
\begin{tabular}{|c|c|c|c|c|c|c|c|}
\hline
Run &Model       
&$\Omega_b$&$\Omega_c$&$\Omega_\Lambda$&$\Omega_\nu$&$n$&$G_{rms}$
\\ \hline
1 & std.  CDM& .05     &   .95   &     --  & --       & 1 &  5.1
\\
2 & std.  CDM& .02     &   .98   &     --  & --       & 1 &  4.7
\\
3 &  tilted. CDM & .05     &   .95   &     --   &--      & .9 &  4.2
\\
4 & Mixed HCDM & .05     &   .75   &     --   & .2    & 1 & 0.9 \\
5 &$\Lambda$-CDM& .05     &   .55   &   .4    & --    & 1 & 2.8 \\
\hline
\end{tabular}}
\caption{Summary of the numerical results. All models are 
flat (curvature free) universes with a Hubble constant of $H_0=66 
km~sec^{-1}Mpc^{-1}$ and with adiabatic initial perturbations. }
\label{result_table}
\end{table}


\section{Can early small scale structure formation be measured?}

We have found that small scale perturbations can under certain
conditions be enormously amplified. The strong enhancement of
particular waves leads to the following questions: First, if the
required conditions do prevail and the amplification factor is more
than enough to drive the perturbation out of the linear regime, at
what point will the growth process stop? Second, if this process did
in fact take place early in our universe, could it have left a mark on
the CMBR that is detectable today? And third, could the process have
had other implications on structure formation in the early universe?


It is difficult to precisely estimate when the growth saturates.  The
simplest estimate to when the growth process ceases is as the wave
reaches the nonlinear regime, a point at which our linear treatment
breaks down anyway.  Clearly, when $\delta\rho/\rho$ is of order
unity, the different speeds of sound at different parts of the wave
can create shocks that will dissipate the waves' acoustic energy into
thermal energy. When $\delta\rho/\rho\sim 1$, the acoustic energy of
isothermal waves is of order the thermal energy of the gas alone (In
adiabatic waves the energy is of order the thermal energy of the gas
and radiation). At the time of recombination, the heat capacity is
very large because a large amount of energy is needed to reionise the
Hydrogen. Thus, after recombination, when the temperature is of order
but less than the recombination temperature $T \lesssim T_{rec}\approx
0.26 eV$, injecting an amount of order the thermal energy of the
plasma will ionise at most only $\sim T_{rec}/E_{ion} \approx 0.26eV /
13.6 eV \approx 1/50$ of the Hydrogen atoms. The free electrons
released by ionization rescatter photons and leave a mark in the CMBR.

Heating the plasma back to the recombination temperature implies that
the plasma and the microwave background are not in equilibrium
anymore. The plasma consequently upsets the Planck distribution of the
radiation through the number conserving Compton scattering
process. The process is evident from Kompaneets' equation (1957) and
its cosmological application by Sunyaev and Zel'dovich (1969). The
dimensionless parameter describing the importance of the effect is the
parameter $y$ defined as:
\begin{equation}
y = {k \left< T_e \right> \over m_e c^2} \int^{t} \sigma_T n_e c dt,
\end{equation}
with standard notations. $n_e$ denotes the number of {\it free}
electrons. The analysis of the COBE data revealed no deviation from a
Planckian spectrum and set a limit of (Fixsen et
al. 1996):\footnote{Note that a smaller upper limit of $y \lesssim
3\times10^{-6}$ (Fixsen et al. 1997) was found when the signal was
correlated with the DMR fluctuation map, however, this type of limit
assumes that $y$ is correlated with the temperature fluctuations, in
our case however, the correlation will be with the temperature
gradients.}
\begin{equation}
y \lesssim 1.5 \times 10^{-5}~~~(95\% CL).
\end{equation}
If the fraction of ionised electrons increases to 
$\iota_e \sim 1/50$, and the
temperature $T$ increases to $\tau T$ with $\tau\sim1$, one can show 
that the resulting $y$ parameter should be of order (e.g. Peebles, 
1993):
\begin{equation}
y \sim 2\times 10^{-11} (\tau-1) (1+z)^{5/2} h\Omega_b
\Omega^{-1/2}\iota_e .
\end{equation}
with $z\sim 1000$ at recombination. Thus, 
\begin{equation}
y_{max} \sim 4 \times 10^{-6} h_{0.75} {\Omega_b}_{0.05} 
\Omega^{-1/2}.
\end{equation}
If through the appropriate nonlinear treatment it would be found that
the kinetic energy of the saturated waves is larger than the thermal
energy then the energy, disspated is larger and $y_{max}$ should be
increased correspondingly. If on the other hand the dissipation is
found to take place on smaller time scales than the universe's age at
the time, $\tau-1$ is smaller and it reduces $y_{max}$. The actual values
cannot be found from our linear analysis.

The aforementioned value for the $y$ parameter is however only in
regions (and directions in the sky) where the gradient at
recombination was large enough to have isothermal waves reach
saturation. The average $y$ will be smaller and given by:
\begin{equation}
y_{avr} = y_{max} {\cal F}\left(\log \delta_0 \over \log G\right).
\end{equation}

Unless the fraction $\cal F$ of the sky that reaches nonlinearity is
of order unity, the $y_{avr}$ is too small to be detectable by present
means. Nevertheless, investigations that correlate $y$ with the
gradients in future sub-degree maps will be able to detect smaller
values for $y$.

 If the observations rule out the existence of such structure then the
 theory imposes strict limits on the cosmological parameters,
 including the fact that the fluctuation distribution function does
 not have a power law tail. However, it can predict black-body
 deformations correlated with the CMBR gradients (i.e., a specific
 correlation). If indeed this is the case, it will be proven as the
 origin for small scale structure, and it could be used for setting
 strict limits on cosmological parameters.

\section{Exotic Possibilities}

More exotic possibilities for the fate of the nonlinear regions rely
on the fact that the sub-Jeans sized objects formed through the
radiative instability during recombination can later on
gravitationally collapse. The typical wave numbers that experience the
large amplification are those with a wavelength larger than the Jeans
scale by at least an order of magnitude. If they exist in the linear
regime they are unstable and collapse when $z\sim 10$, however, these
wave lengthes are already nonlinear at $z\sim 1000$ and may easily
collide and collapse long before their linear cousins will. Compact
objects of order of the Jeans scale can therefore form and collapse at
very high redshifts $10 \lesssim z \lesssim 1000$. What will the fate
of these objects be?  They might collapse into massive Black holes,
they can also burn and release a few$\times 10^{-5}$ of their energy
while doing so (Bond et al. 1984). These nonlinearities may perhaps
form globular cluster like clusters or maybe even seeds for AGN
engines.

Another possibility is that the nonlinear objects form structure on
larger scales through the large energy they release and a chain
reaction of explosions that they ignite (Ostriker \& Cowie, 1981),
they can then prompt the collapse of galaxies in the expanding shells
and form voids. Although the peculiar velocity and the $y$-distortion
prediced for the simplest of these models are both too high (Levin,
Freese \& Spergel, 1992), the inclusion of continuous or multi-cycled
detonation waves in the explosion model can lead to large voids with
a diameter larger than 10 Mpc, a peculiar velocity of roughly
$100~km~sec^{-1}$ and a $y$-distortion less than the limit set by COBE
(Miranda \& Opher, 1997). If these explosions can indeed be ignited,
only a very small fraction of the universe needs to reach the
nonlinear regime by the end of recombination in order to have
dramatic effects on the formation of large scale structure.

\section{Feedback on larger scales}

The next question to address is whether the effect can also affect the
large scales. Since energy is conserved, large scales will be affected
if the energy transferred to the smaller scales is comparable to the
original energy in the large scales.  The energy per unit mass in the
large scale waves is roughly given by $(\delta \rho_{ad} / \rho)^2
v_{(a)}^2/2$ while the energy per unit mass in the small scale
isothermal waves is roughly $(\delta \rho_{iso} /
\rho)^2 v_{(i)}^2/2$. Inspection of eq. \ref{eq_14}  reveals
that when the small isothermal waves have a dimensionless amplitude of
order unity over a large fraction of the universe and $G_{rms}
\sim 1$, the energy content in both types of waves is comparable. A
significant fraction of the energy in the large scales can therefore
be transferred to the small scales.

 Apparently, an exact treatment is beyond the scope of the linear
 analysis developed here and the unknown nonlinear physics can at most
 be phenomenalised using few parameters and a simple demonstrative
 model.

Using the terminology and approximations of \S2, one can express the
original energy per unit mass in the large scales as $A_l^2 G_{rms}^2
\times v_{(i)}^2/2$. Under the approximation of \S2, we have  $A_l=4$,
however, for optically thin large scale waves the energy in the waves
is different and $A_l$ can change.  The energy in the isothermal waves
is $A_s^2\times v_{(i)}^2/2$ and the energy growth rate is $2 G_{rms} A_s^2 
/\Delta t  \times v_{(i)}^2/2$. 

Next, we assume the energy damping rate of the small scale waves to be
of the form:
\begin{equation}
\left. {d E\over dt}\right|_{damping} \equiv {A_d\over \Delta t}
A_s^\epsilon \times v_{(i)}^2/2.  
\end{equation}
$\Delta t$ is the duration of recombination.  $A_d$ and $\epsilon$ are
parameters that represent the unknown physics. Both parameters can
depend on the wavelength of the unstable isothermal waves. $A_d$
should be of order unity for waves with a period of $\sim \Delta t$,
but otherwise unknown. The power $\epsilon$ with which the damping
depends on the amplitude of the isothermal wave is also unknown but
should satisfy $\epsilon > 2$ if it is characterised with only one
power index and if it is to be more important on the nonlinear scales
than on the linear ones.

If the value of $G_{rms}$ before the instability commences is large
enough then a large fraction of the volume leaves the nonlinear regime.
The critical value for this to take place is
\begin{equation}
G_{rms} \gtrsim -\ln (\delta \rho_{iso}/\rho).
\label{nonlinearity_eq}
\end{equation}
The simplest simplifying assumption is that the isothermal waves will
saturate at a level where the energy growth rate is equal to the
dissipation rate.  The small scale amplitude $A_s$ will then be given 
by
\begin{equation}
A_s = \left( G_{rms} \over A_d\right)^{1\over \epsilon-2}.
\end{equation}
 Compare now the energy in the large scales with the total energy that
 can be dissipated.  When the latter energy is larger, all the energy
 in the large scales can be dissipated.  The condition for it to take
 place is:
\begin{equation}
\left\{
\begin{array}{c l}
G_{rms}  \lesssim  2^{\epsilon-2} A_d^{2 /(4-\epsilon)}
A_l^{2/ (4-\epsilon) }  &{\rm for} ~\epsilon<2~{\rm or}~\epsilon>4 \\
G_{rms}  \gtrsim  2^{\epsilon-2} A_d^{2 /( 4-\epsilon)}
A_l^{2/( 4-\epsilon) }  &{\rm for}~2<\epsilon<4
\end{array}
\right.
\label{damping_cases}
\end{equation}
When $2<\epsilon<4$, an initial value of $G_{rms}$ that satisfies the
above condition will imply that the amplitude of the large scale
fluctuation can be reduced until $G_{rms}$ does not comply with the
above condition. For other values of $\epsilon$, the initial value of
$G_{rms}$ must be smaller than a critical value in order to satisfy
the above condition, in which case, the large scale amplitude will
apparently be reduced arbitrarily. In fact, $G_{rms}$ continues to
decrease until the shear drops below the critical value separating
between the two cases, at which moment the instability turns
off. Since $\tau_{e\gamma} \sim 1$ at the end of recombination (see
e.g. figure
\ref{runput4_figure}), the switching off will take place at $G_{rms}
\sim 1$.  To summarise, two possibilities for the damping of large
scale waves seem to exist depending on the power $\epsilon$ relating
the dependance of the damping rate with the amplitude of the
isothermal wave.  In the first case, any $G_{rms}$ larger than roughly
the critical value given by eq.~\ref{damping_cases} will be reduced to
a value of order of critical value provided that it satisfied
eq.~\ref{nonlinearity_eq} as well. In the second case, any $G_{rms}$
which satisfies both eq.~\ref{damping_cases} and
eq.~\ref{nonlinearity_eq}, will be reduced to $G_{rms} \sim 1$.

\begin{figure*}
\centerline{\epsfig{file=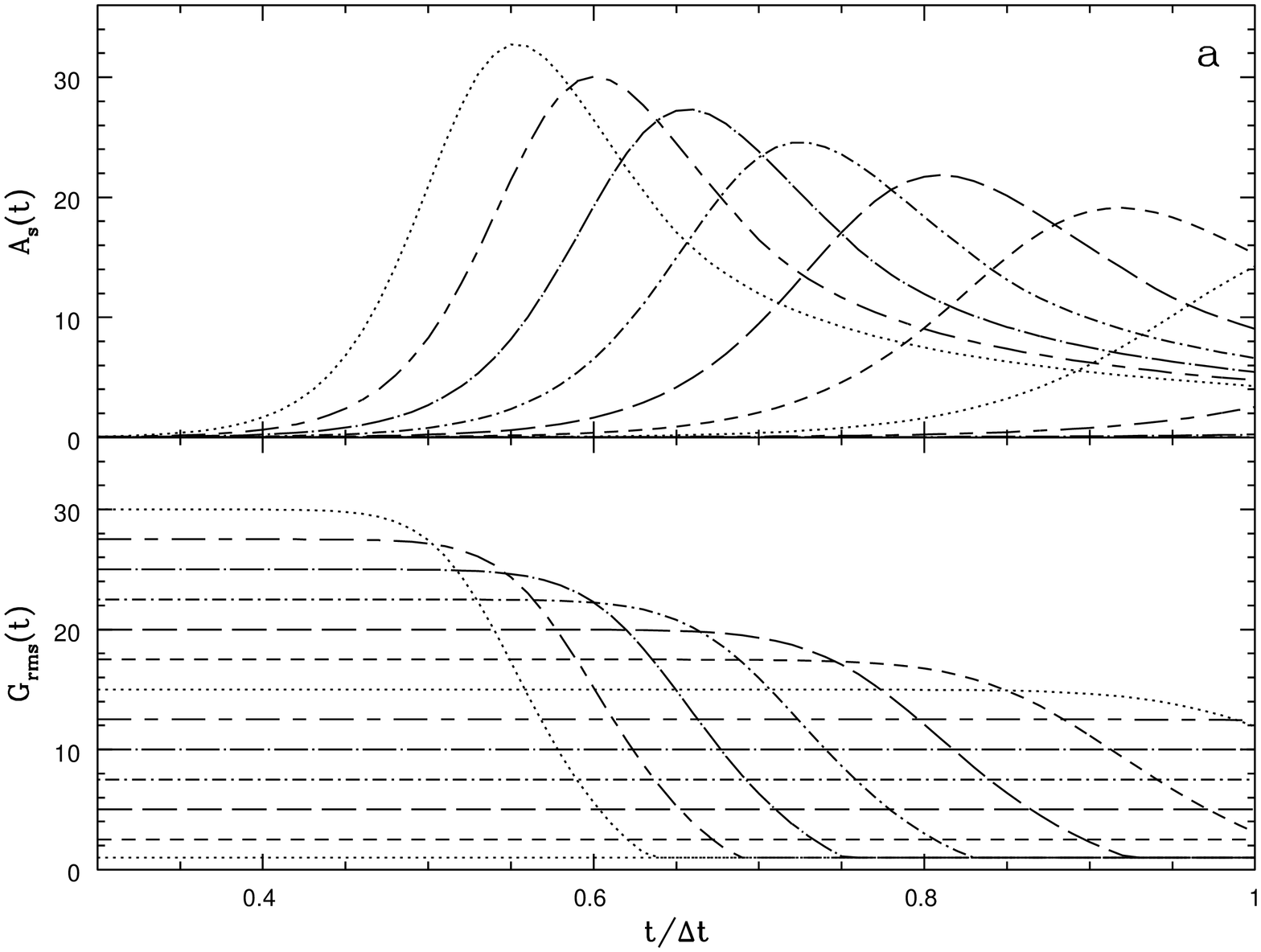,width=\figurewidth,angle=-90}}
\vskip -10pt
\centerline{\epsfig{file=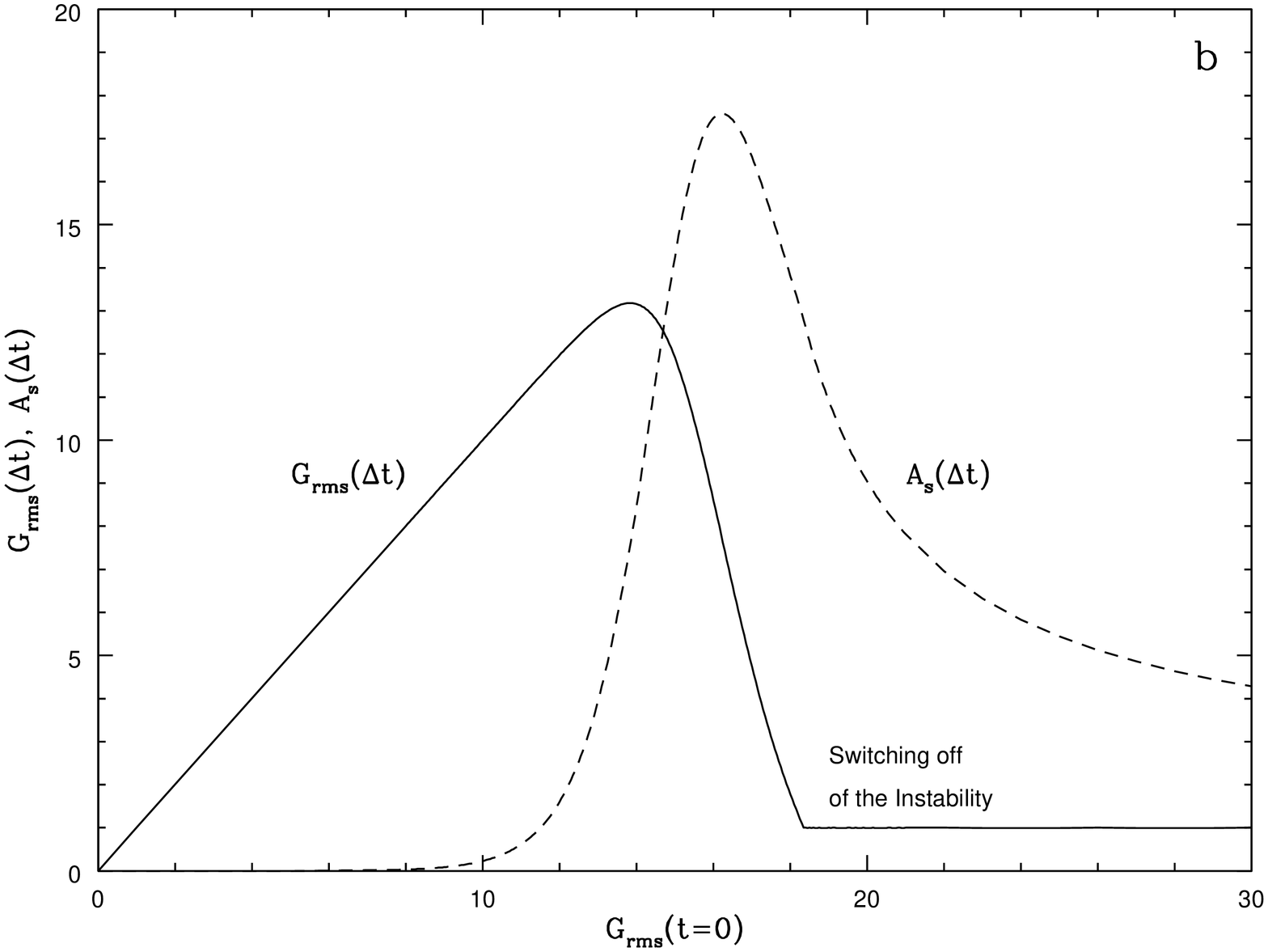,width=\figurewidth,angle=-90}}
\caption{
The two panels represent the results of the numerical integration of
the two approximate differential equations describing the evolution of
the energy in the large scales ($A_l^2 G_{rms}^2 \times v_{(i)}^2/2$)
and in the small scales ($ A_s^2 \times v_{(i)}^2/2$). The initial
small scale amplitude before the instability takes place is
$A_s(t=0)=10^{-5}$. Also, $\epsilon=3$, $A_d=1$ and $A_l=4$. The
evolution of $A_s(t)$ and $G_{rms}(t)$ is depicted in panel a for
various initial values of $G_{rms}$. Note that only part of the
recombination period $\{0,\Delta t\}$ is shown. The same cases are
drawn with the same line type in both parts. Panel b depicts the final
values at the end of recombination -- $A_s(t=\Delta t)$ and
$G_{rms}(t=\Delta t)$. For the switching between cases A \& B of table
1, it was assumed that the critical shear corresponds to $G_{rms}
=1$. It can however be somewhat larger or smaller.}
\label{dissipate_figure}
\end{figure*}

As an elucidation, one can numerically solve a simplified toy model
for the process. The transfer of energy from the large scales into the
small scales and then the dissipation by small scale nonlinearities
can be simplified and described with the two differential equations:
\begin{eqnarray}
 {d\over dt} \left( A_l^2 G_{rms}(t)^2 \right) & = & - {2 
G_{rms}(t)\over \Delta T} A_s(t)^2 \\
 {d\over dt} \left( A_s(t)^2 \right) & = & + {2 G_{rms}(t)\over 
\Delta T} A_s(t)^2 - {A_d\over \Delta T}
A_s(t)^\epsilon. \nonumber
\end{eqnarray}
Figs.~\ref{dissipate_figure}a \& \ref{dissipate_figure}b describe the
numerical results for the integration of the two differential
equations when $\epsilon=3$, $A_d=1$, $A_l=4$, for initial
fluctuations of $\delta\rho_{iso}/\rho=10^{-5}$ and different initial
values of $G_{rms}$. By comparing to the numerical results to
eq.~\ref{damping_cases}, it is clear that the analytical estimate is
an over simplification of the differential equations, which too are a
simplification of the real physics. Nevertheless, the main conclusion
is that for a large range of an initial large-scale amplitudes, it is
possible to have almost all the energy dissipated into smaller scales
and subsequently have it dissipated through nonlinear processes. The
large scales then reach their natural lower limit -- the value for
which the large scale shear switches off the instability.

The implications are very interesting.  The measured fluctuations in
the microwave background radiation reflect the large scale structure
after the fluctuations were reprocessed by the instability during
recombination.  In other words, the value oberved does not necessarily
reflect the pre-recombination value of the fluctuations. It is
possible to have an initially large large-scale amplitude, induce
nonlinear small scale structure and eventually reduce the large scale
amplitude.

\section{Discussion}

We have analysed the nonlinear growth of small scale isothermal
fluctuation through the induced effect that large-scale perturbations
have. The equations are solved through the separation of the large
scale inhomogeneities into smaller regions where they can be
considered uniform, and the shear $\Delta v_0$ exerted by the large
scales is constant in space.  This can be performed because the waves
that interest us are waves with very small wavelengths and a very
small propagation speed, such that over the integration time, the
waves practically remain in the same region of space.

Small scale isothermal waves are found to be unstable if the large
scale shear velocity between the ``photon fluid'' and the baryon fluid
is larger than roughly twice the isothermal speed of sound at the time
of recombination. A period when the logarithmic derivative of the
opacity with respect to the density does not vanish but is $-1/2$
instead.  When the shear is smaller than the critical value, the
solution found is qualitatively the same as the solution for the
completely decoupled equations. The shear in this case may induce some
quantitative corrections but the behaviour of the system does not
change.  The behaviour of the system is however radically different if
the shear is larger than a critical value.  For wavelength shorter
than the roughly the Jeans scale, the growth rate increases with $k$
and it saturates when $\tau_{e\gamma}=\tau_o$, i.e., at a scale which
is a few times smaller than the Jeans scale. The maximal growth rate
attained is:
\begin{equation}
r_{\infty} = {\alpha \over 2} {\Delta v_0 \over v_s} {1\over
\tau_e{\gamma}}.
\end{equation}
 If $r_{\infty}$ is much larger than all other rates in the system, it
 can be easily integrated over the duration of decoupling $\Delta t$.
 The growth factor obtained is then:
\begin{equation}
{\delta_1\over\delta_0} \approx \exp (G_{\infty}) = \exp
\left(\left\langle {\alpha \over 2} {\Delta v_0 \over v_s}
{1\over \tau_e{\gamma}}\right\rangle_t \Delta t\right),
\end{equation}
with the brackets denoting a temporal average. The isothermal waves
are unstable because their sound speed is very small. Consequently,
the acoustic energy in a wave with a given amplitude is very small and
the amount of work needed to increase its amplitude isn't large, in
fact, it is so small that nonlinear effects start to take place
already at $\delta\rho/\rho\sim 10^{-5}~{\rm to}~10^{-4}$!

Even more surprising is the fact the the typical shears found in
typical cosmological scenarios is {\em exactly} at the verge of having
nonlinear effects take place! If the amplitude of the large scale
fluctuations would have been an order of magnitude larger, then any
isothermal perturbations, however small, could have been amplified
right into the nonlinear regime. On the other hand, if the typical
amplitude would have been a few times smaller, the effect would have
been completely meaningless.

For typical cosmological scenarios normalized to COBE, the growth
parameter is roughly $G_{rms} \sim 1 - 5$, e.g., for an
$\Omega=1,~\Omega_b=0.05,~h=.66$ CDM model one finds $G_{rms} \approx
5$, if the spectral index is tilted to $n=0.9$, one finds
$G_{rms}\approx 4$, if a hot component is added at a 20\% level to the
untilted model, it falls to $G_{rms} \sim 1$.

The fate of the regions that do reach nonlinearity is still an open
question. Do these regions collapse and form black holes?  Do they
form an early generation of massive stars? Do they release enough
energy to the environment and affect it, or perhaps, even form
structure on much larger scales? Although some speculations of what
might occur do exist in the context of early small scale structure
formation, it is not entirely clear. Moreover, nonlinear hydrodynamic
simulations is probably unavoidable if we are to really solve the
problem. The only relatively certain consequence is that the
dissipation taking place after recombination heats the matter and it
is likely to ionise it, raise its temperature to the ionisation
temperature and leave an imprint on the CMBR in the form of a
deviation from a Planckian spectrum. In scenarios where
$G_{rms}\gtrsim 5$, isothermal perturbations with an amplitude of as
low as $\delta\rho_{iso}/\rho\sim10^{-6}$ will be in fact
theoretically detectable in the future. An amplitude of
$\delta\rho_{iso}/\rho\sim10^{-4}$ will leave 1\% of the background
sky with a detectable deviation. This fraction can change considerably
if the density perturbations are not Gaussian.

The detection of deviations from a Planckian spectrum or the placing
of an upper limit for them can result with interesting
implications. Any deviations found will first of all directly prove
the existence of primordial isothermal fluctuations. Second, such a
detection will place stringent limits on cosmological parameters, as
only those scenarios that produce a large enough $G_{rms}$ are capable
of producing nonlinear structure, and because the amount of
nonlinearity is extremely sensitive to $G_{rms}$, it is also sensitive
to the exact cosmological parameters. Third, if the cosmological
parameters are known with an large accuracy (for example, through the
fitting of the CMBR spectrum) then the primordial isothermal
perturbation spectrum at very large $k$'s can be estimated.

Even if no detection of a deviation from a Planckian spectrum is
found, one can place limits on cosmological parameters. Moreover, if
in the future it will be found that the cosmological parameters
actually correspond to a high $G_{rms}$ model, one will be able
through the lack of detection of a $y$-parameter to place extremely
powerful limits on the amplitude of the isothermal component of the
primordial spectrum. Through the Press \& Vishniac effect, limits can
also be placed on the adiabatic spectrum.

Another interesting implication is the possibility of transferring
enough energy from the large to the small scales and considerably
change the amplitude of the large-scales. This possibility relies on
whether a significant volume fraction can be amplified out of the
linear regime and on the maximum dissipation rate of nonlinear
isothermal waves. Under favourable circumstances, the large scale
amplitude can be significantly reduced such that $G_{rms}$ calculated
from the observed large scale fluctuations would only be of order
unity. The fact that the observed fluctuations in the CMBR correspond
to values of this order raises a very interesting question.  Why do
the values of $G_{rms}$ corresponding to the observed fluctuations
happen to fall in a small region around unity? Is it because the
universe was created with a fluctuation spectrum corresponding to this
region, or, is it because $G_{rms}$ that evolved from the initial
cosmological parameters was actually larger but it was naturally
reduced to the oberved value?

One can summarise the several plausible scenarios according to both
the value of $G_{rms}$ predicted by the cosmological model and the
amplitude $\delta = \delta \rho_{iso}/\rho \approx 10^{-10} - 10^{-5}
$ of the sub Jeans scale isothermal fluctuations before the advent of
recombination, these possibilities are:

\begin{enumerate}

\item
If $G_{rms}$ is of order unity or less, the typical growth of
isothermal waves is at most a few $e$-folds. The effect will be
insignificant as the predicted micro degree size fluctuations are
neither fluctuations of the temperature nor on a scale measurable in
the near future. Moreover, there are no implications at all to
structure formation.

\item
 For a value of $G_{rms} \gtrsim - (\ln \delta)/3 \approx 4 - 8$, the
 effect will be measurable as patches in the CMBR with a distorted
 Planckian spectrum. As long as $G_{rms} \lesssim - \ln \delta \approx
 10-20$, only a small fraction of the universe would have reached
 nonlinearity and formed small scale structure by the end of
 recombination. Under certain circumstances however, it can lead to
 large scale structure formation through explosive amplification.

\item
 For a value of $G_{rms}$ such that $ G_{rms} \gtrsim - \ln \delta
 \approx 10-20$, a large fraction of the universe reaches nonlinearity
 by the end of recombination and small scale structure is subsequently
 formed in most of the universe soon after recombination. The large
 scale spectrum is damped by transferring a significant amount of
 energy to the small scales, thus, the $G_{rms}$ measured from the
 damped spectrum is smaller than the $G_{rms}$ calculated when
 neglecting this process. In some cases, $G_{rms}$ will be reduced to
 a values of order unity and it will mimic values on the boundary
 between the first and second scenarios. Note that it will not affect
 fluctuations in components that decouple from the matter-radiation
 fluid before recombination (e.g. dark matter fluctuation).

\end{enumerate}

The analysis presented here is the first step in the investigation of
the radiation-matter interaction instability. More accurate
integration is needed to improve the actual transfer function for
isothermal waves.  The analysis here was restricted to the evaluation
of $G_{rms}$ and it does not include the actual rates which are $k$
dependent. This approximation overestimates the rate of growth of
finite sized $k$'s.  In the present paper we have simulated the
freezing of the instability due the freezing of recombination by
stopping the integration abruptly at a given $z$. However, the
equilibrium conditions and therefore the switching off of the effect
depend on the isothermal wavelength as well.  By assuming this
assumption, we have actually underestimated the contribution from waves
of order the Jeans size.

Many of the possible implications depend on the quantitative behaviour
of of the nonlinearities once they are reached. A numerical
hydrodynamic study of large amplitude waves will certainly help us
understand of the fate of the nonlinear objects and the possible
implications they have on large scale structure as well.

\section*{Acknowledgements}
The author wishes to thank Yuri Levin for the fruitful discussions and
Caltech for the DuBridge Prize Fellowship supporting him. The author
is also grateful for the readily available COSMICS code written by
Bertschinger and Ma (under NSF grant AST-9318185).

\end{document}